\documentclass[tradiabstract]{aa}
\usepackage{newtxtext,newtxmath}
\usepackage[T1]{fontenc}
\usepackage{xcolor,ulem}

\usepackage[colorlinks=true]{hyperref}
\hypersetup{citecolor=blue,linkcolor=magenta}
\usepackage{graphicx}
\usepackage{color}

\definecolor{maroon}{cmyk}{0,0.87,0.68,0.32}

\usepackage{amsmath,graphicx}
\newcommand{\JT}{\tau_{\text{jet}}}
\newcommand{\spin}{\nu_\star}

\usepackage{journals}
\begin{document}

\titlerunning{Measuring NS spin via double precursors}
   \title{Measuring spin in coalescing binaries of neutron stars showing double precursors}

   \author{Hao-Jui Kuan\inst{1,2,3} \and
   Arthur G. Suvorov\inst{1,4}
          \and
          Kostas D. Kokkotas\inst{1}
          }

   \institute{Theoretical Astrophysics, IAAT, University of T{\"u}bingen, T{\"u}bingen, D-72076, Germany
         \and
             Max Planck Institute for Gravitational Physics (Albert Einstein Institute), 14476 Potsdam, Germany
             \and
             Department of Physics, National Tsing Hua University, Hsinchu 300, Taiwan
             \and
             Manly Astrophysics, 15/41-42 East Esplanade, Manly, NSW 2095, Australia
             }

   \date{Received ???; accepted ???}

\label{firstpage}

\abstract
   {Gamma-ray bursts resulting from binary neutron-star mergers are sometimes preceded by precursor flares. These harbingers may be ignited by quasi-normal modes, excited by orbital resonances, shattering the stellar crust of one of the inspiralling stars up to $\gtrsim10$ seconds before coalescence. In the rare case that a system displays two precursors, successive overtones of either interface- or $g$-modes may be responsible for the overstrainings. Since the free-mode frequencies of these overtones have an almost constant ratio, and the inertial-frame frequencies for rotating stars are shifted relative to static ones, the spin frequency of the flaring component can be constrained as a function of the equation of state, the binary mass ratio, the mode quantum numbers, and the spin-orbit misalignment angle. As a demonstration of the method, we find that the precursors of GRB090510 hint at a spin frequency range of $2 \lesssim \nu_{\star}/\text{Hz} \lesssim 20$ for the shattering star if we allow for an arbitrary misalignment angle, assuming $\ell=2$ $g$-modes account for the events. }

\keywords{
gamma-ray burst: individual: 090510 -- stars: neutron, oscillations -- gravitational waves
}

\maketitle

\section{Introduction}
 
Some short gamma-ray bursts (SGRBs), which are thought to originate from binary neutron-star (NS) mergers, are preceded by precursor flares with a time advance that ranges from $\sim 1$ to $\gtrsim 10$ s \citep{Troja:2010zm,Minaev:2018prq,Zhong:2019shb,Wang:2020vvr}. 
These early flashes may be caused by crust yielding in magnetised NS members, resulting from resonantly excited quasi-normal modes (QNMs) \citep{Tsang:2011ad,Tsang:2013mca,Suvorov:2020tmk,Kuan:2021sin}. In this context, the timing of a precursor relative to the SGRB, which also depends on a jet formation/breakout timescale, estimates the frequency of the mode that leads to the crustal fracture. On rare occasions, more than one precursor precedes the SGRB, for which the frequencies of the two responsible modes may be acquired \cite[e.g.,][]{Kuan:2021sin}. 

Certain details of the stellar fabric can be accessed from the QNM spectrum, e.g., the interior mean density strongly correlates to frequencies of pressure modes \citep{Andersson:1997rn,Kruger:2019zuz}, and $g$-modes encode microphysical temperature or composition gradients.
Here we discuss a novel way to learn the spin of a NS if a double precursor event is observed. In particular, mode frequencies in a rotating NS, attributable to the pre-emissions, provide two relations between the free mode frequencies of these two modes and the stellar spin. In scenarios where the free mode frequencies have a constant ratio, such as for $g$- and $i$-modes as explained below, this additional relation then allows for a spin inference. In the current era of gravitational-wave (GW) astrophysics, estimating the spins of binary NSs is crucial in shrinking the error in other measurements \citep[e.g.,][]{ma21,gupt23}; for instance, the estimates of tidal deformability of GW170817 and GW190425 are sensitive to the spin priors assumed for the progenitors \citep{gw170817,ann18,lvk_prx,lvk_cqg}. Properties of the post-merger system, such as the gravitational waveform \citep{kast17}, the content of dynamically ejected matter \citep{fuji18}, remnant disc mass \citep{east19}, and the kilonova \citep{papan22} also depend sensitively on the spins of the pre-merging stars.
In addition, simultaneous knowledge of the spin and the mode frequencies may set strong constraints on the equation of state \cite[EOS;][]{Biswas:2020xna}. A particular example to manoeuvre out the spin is implemented for SGRB 090510 in this work, an event preceded by two precursors occurring $\sim13$ and $\sim 0.5$ s prior to the main burst \citep{abdo09,Troja:2010zm}.

Section \ref{sec:precursors} of this article briefly reviews the theory of resonant shattering as a mechanism for precursor ignition, with an emphasis on $g$-modes (though see also Sec. \ref{sec:diff}). Theoretical predictions based on binary formation channels are considered in Sec. \ref{sec:form}, as relevant for misalignment angles and timing considerations (Sec. \ref{Sec:tilted}). Section \ref{Sec:spinfreqdet} forms the main part of the paper, and demonstrates, in principle, how the timing of double precursors can constrain the spin frequency of the flarer. {By exploiting the approximately constant ratio between $g_1$- and $g_2$-modes, and how stellar spin modifies the mode frequencies, our key result is a fitting formula [Eq.~\eqref{eq:spin}], that takes into account tidal heating between fracture events (Sec.~\ref{tidheat}), for estimating the spin of the star assuming that it emits two precursors. Some discussion on uncertainties due to a jet formation/breakout timescale is also covered in Sec.~\ref{sec:jet} for completeness}. Some discussion on Blue/red kilonovae and GWs from the remnant is offered in Sec. \ref{sec:otherobs}. The article is summarised in Sec. \ref{sec:summary}.

\section{GRB precursors via \textit{g}-mode resonances}
\label{sec:precursors}

Although the definition of pre-emission in SGRBs is not uniquely given as, for instance, some authors require the waiting time to be longer than the main burst duration \citep{Minaev:2018prq} while others do not \citep{Zhong:2019shb}, precursor flares have been confidently identified in rare ($\lesssim 10\%$) cases \citep{Wang:2020vvr}. These early flares may be triggered by certain, resonantly-excited QNMs \citep{Tsang:2011ad,Suvorov:2020tmk,Kuan:2021sin}. The (linear) orbital frequencies of precursors, which are uncertain owing to a delay between the main GRB and the merger through a jet formation and subsequent breakout timescale, suggest that $\sim 100$~Hz modes are promising to account for the pre-emissions; in particular, shear-, interface-, and $g$-modes have attracted some attention \citep{Tsang:2011ad,Tsang:2013mca,Kuan:2021sin}. We focus on the $g$-mode scenario in this article since we may accommodate double precursors by one class of modes, though a discussion about other modes is given in Sec. \ref{sec:diff}.

 \subsection{Parameterised \textit{g}-modes}\label{reso}

Composition and/or temperature gradients stratify the interior of a NS, so that it may support $g$-modes. The spectrum of these modes {is determined by the `adiabatic index' of the fluid perturbation relative to that of the (beta-equilibrium) background, setting the characteristic Brunt-V{\"a}is{\"a}l{\"a} frequency \cite[e.g.,][]{reis01}. In general,} the index depends on the respective Fermi energies of each particle species, {most notably through the electron fraction $Y_{e}$, and the temperature of the star} \cite[e.g.,][]{haen02}. 
Realistic profiles for these quantities are complicated, and depend {on a number of largely-uncertain aspects of the stellar interior} \cite[see][for a review]{latt12}. 
In this work, our main goal is to illustrate a method by which spin can be measured in NSNS mergers that release two precursor flares. To this end, we work within the context of the simple, toy framework described by \cite{Kuan:2021jmk} \cite[see also][]{Gaertig:2009rr,pass09,xu17,Passamonti:2020fur}, where stratification is encoded in a spatially-constant but time-dependent parameter $\delta$, defined as the difference between the (generally density-dependent) adiabatic indices of the perturbation and the background star\footnote{{Using the introduced stratification parameterization we find, for a family of WFF EOS, the $g_1$-mode frequency is $\sim90$Hz for a NS with $M_\star=1.4M_\odot$ and $\delta=0.005$. This matches to the self-consistently obtained frequencies of \cite{lai94} to within $20\%$ [e.g., for the EOS we call WFF1 but they call `AU', we find $93.03$ ($61.11$) Hz while they get $72.6$ ($51.4$) Hz for the $g_1$- ($g_2$-)mode frequency]. Note also that \cite{lai94} uses a Newtonian scheme while ours is general-relativistic, likely accounting for most of the disparity. The validity of the spatially-constant $\delta$ approximation specifically is detailed in Appendix \ref{sec:app}.}}. More specifically, we define\footnote{This expression differs slightly from Equation (10) in \cite{prep} due to a typographical error in that work. Nonetheless, the results therein are essentially unaffected: $g$-spectra with spatially-varying $\delta$ were in fact computed in \cite{prep}, where it was concluded that (i) the $g$-spectrum is largely determined by the surface temperature since only in the outer most part of the star can buoyancy be comparable to the isotropic pressure, and (ii) the constant $\delta$ approximation works well for surface temperatures below $\sim 10^{10}$~K.}
\begin{equation}\label{eq:deltadef}
	\delta(t,\boldsymbol{x}) =\left[ \frac{k^2\pi^2}{6}\sum_x \frac{n_x(\boldsymbol{x})}{E_F^x(\boldsymbol{x})} \right] \frac{T(t,\boldsymbol{x})^2}{p(t,\boldsymbol{x})},
\end{equation}
for pressure $p$ and temperature $T$, where particle species $x$ has number density $n_{x}$ and Fermi energy $E^{x}_{F}$, and the sum runs over the species list (treated as being just non-relativistic $n$ and $p$, for simplicity), and assume $\nabla_{j} \delta \approx 0$ (see Appendix \ref{sec:app}). In the time between $\sim 10^{2}$~s before the first precursor and the merger, the composition of the star changes very little, though the temperature can evolve dramatically \cite[e.g.,][]{lai94}. As such, both thermal and compositional gradients define $\delta(t_{0},\boldsymbol{x})$ for simulation start time $t_{0}$, while only thermal gradients then contribute to the \emph{evolution} of $\delta$ due to tidal effects and mode-induced backreaction (see Sec. \ref{tidheat} for more details). The shift in $g$-mode spectra at late times is largely attributable to heating, even though the composition gradient is the main source of stratification; some studies suggest an effective $\delta\gtrsim0.01$ for compositional stratification \citep[e.g.,][]{reis09,akgu13}.

It should be recognised therefore that the numerical estimates we provide for spin frequencies are subject to some systematic uncertainty, and do not necessarily represent realistic, astrophysical predictions. In principle, however, one could solve the relevant thermodynamic system, for a given EOS, to self-consistently identify the value of $\delta$ for a particular type of perturbation and (magneto-)hydrodynamic equilibrium. Such complications include the possible existence of superfluidity and/or a hadron-quark transition in the core, both of which lead to larger $g$-mode frequencies \cite[e.g., ][]{Yu:2016ltf,Jaikumar:2021jbw}. We endeavour to present formulae in such a way that the reader can readily substitute alternative values, to pave the way for more realistic investigations in future.

\begin{figure}
	\centering
	\includegraphics[width=\columnwidth]{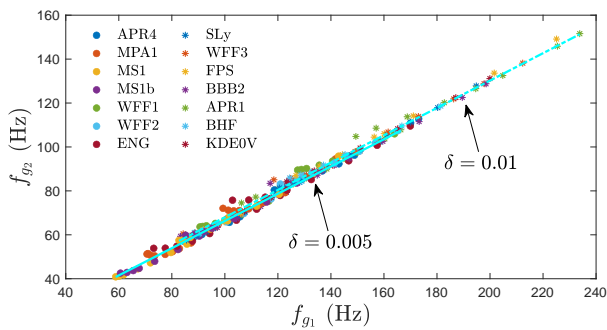}
	\caption{Correlation between the frequencies of $g_1$- and $g_2$-modes for various EOS and two representative values of $\delta$ (see plot legends). For a given $\delta$, the relations between $g_1$- and $g_2$-modes [Eq.~\eqref{eq:g1g2}] are shown as bright blue lines with the solid one applying for $\delta=0.005$ and the dash-dot one applying for $\delta=0.01$. }
	\label{fig:g1g2}
\end{figure}

{In the context of $g$-modes, the restoring force is much weaker than that supplied by the hydrostatic pressure in the NS core, meaning that $g$-mode motions are suppressed in this region. The stratification in the crust then largely determines the $g$-mode spectrum. This was confirmed numerically in \cite{prep} for relativistic stars, who found quantitatively similar spectra for various spatially-varying $\delta(\boldsymbol{x})$ profiles relative to cases with constant $\delta$, as long as the surface values match (see Sec.~2.1 therein and Appendix \ref{sec:app}). Either way, most EOS predict a typical value} for mature NSs of $\delta\gtrsim0.005$ \citep{xu17}, while the free mode frequencies of $g_1$- ($f_{g_1,0}$) and $g_2$-modes ($f_{g_2,0}$) scale as the square root of $\delta$, viz., $f_0=\alpha \sqrt{\delta}$ for a parameter $\alpha$ depending on EOS, mode's quantum-number, and the mass ($M_{\star}$) and radius ($R_{\star}$) of the NS \citep{prep}. In addition, they can be related via (see Fig.~\ref{fig:g1g2})
\begin{align}
    f_{g_2,0} = 0.62f_{g_1,0} + \beta,    \label{eq:g1g2}
\end{align}
for an EOS-independent parameter $\beta \approx 4.32$ Hz. {[Note that, in the high-$n$ limit, the ratio $f_{g _{n+1}}/f_{g_{n}}$ becomes exactly $n/(1+n)$ for any EOS \cite[see expression (5.13) in][]{lai94}]}. However, QNM frequencies of forced systems deviate from those of free systems; given a perturbing force $\delta\boldsymbol{F}^{\mu}$, a frequency shift of \citep{unno,Suvorov:2020tmk}
\begin{align} \label{eq:unno}
    \delta f=  \frac{1}{8\pi^2f_{0}}
	\frac{{\int\delta \boldsymbol{F}_{\mu}\overline{\xi}^{\mu} \sqrt{-g}d^{3}x}}
	{{\int(\rho+p)e^{-2\Phi}\xi^{\mu}\overline{\xi}_{\mu}\sqrt{-g}d^{3}x}},
\end{align}
is induced for Lagrangian displacement $\xi^{\mu}$, free-mode frequency $f_{0}$, mass density $\rho$, lapse function $\Phi$, and metric determinant $g$.
For centrifugal forces in stars rigidly rotating at a rate of $\spin$, we have a simple expression \cite[see~Eqs.~(70) and (71) of][]{Kuan:2021jmk}
\begin{align}\label{eq:rotshift}
    f=f_0-m(1-C)\spin,
\end{align}
for the \emph{inertial-frame} frequency $f$. The constant $C$ depends on EOS and the mode quantum numbers, the azimuthal one of which, $m$, leads to a Zeeman-like splitting of the modes \cite[e.g.,][]{Kruger:2019zuz}.
As shown by \cite{Kuan:2021jmk}, rotation affects the frequencies of $g_1$- and $g_2$-modes to a similar extent; denoting the constant in Eq.~\eqref{eq:rotshift} for $g_1$- and $g_2$-modes by, respectively, $C_1$ and $C_2$, for a broad set of EOS (see the legend of Fig.~\ref{fig:g1g2}), we find two facts about $C_1$ and $C_2$ for $10^{-3}\le\delta\le0.05$: (i) Both depend only weakly on $M_{\star}$ and EOS. (ii) The maximum difference between the value of $C_1$ and $C_2$ is $\sim 13\%$. (iii) The values of both are 0.11--0.12, while we note that the Newtonian approximation yields $C\simeq 1/\ell(\ell+1)=0.167$ for $\ell=2$ \citep[see, e.g.,][]{vavo08}.

In addition to spin-induced modulations, tidal \citep{Kuan:2021jmk,yu23} and redshift \citep{stei16,zhou23} factors also influence the mode frequency. The former effect can be taken into account by including the tidal `force' within Eq.~\eqref{eq:unno}, though the resulting shift in $g$-mode frequency is less than the spin-induced one by at least 3 orders of magnitude for $\nu_{\star} \gtrsim 1\,$Hz [see Sec.~5.2 of \cite{Kuan:2021jmk}]. Redshift effects (that is, accounting for the fact that frequencies differ between the neutron star and laboratory frames) are more important, though still relatively small. Assuming a circular orbit and an equal-mass binary ($q \approx 1$), the relevant redshift factor is encoded in the lapse function which, to leading (post-Newtonian) order and ignoring spin corrections, reads \citep[cf. Eq. (3.6) of][]{stei16}
\begin{equation} \label{eq:redshift}
    z \approx 1 - \frac{5 G M_{c}}{4ac^2} = 1 - 0.03 \left(\frac{M_{c}}{1.6 M_{\odot}}\right) \left(\frac{100\,\text{km}}{a}\right).
\end{equation}
Given that the onset of $g$-mode resonances are typically $\sim$1--10~s before merger, the separation at resonance is $a\gtrsim100$ km. As expression \eqref{eq:redshift} shows, we expect a frequency shift of at most a few per-cent and so  neglect the effect here. By contrast, the redshift is sizeable for $f$-mode resonances \citep{stei16}.

\subsection{Resonant shattering}\label{shattering}
As the NSs inspiral, tidal fields induce perturbing forces that act predominantly at frequencies that are twice the orbital frequency \citep{zahn77,Tsang:2011ad}. The forced system for QNMs suggests that a particular mode will be brought into resonance when its frequency matches the forcing rate, where the mode amplitude increases rapidly until hitting a ceiling value that depends on the so-called overlap integral. After the mode leaves the resonance window, its amplitude decays by viscosity. The timescale of the dissipation is generally much longer than the rest of the life of the coalescing binary, whereas the resulted heating may change the stellar condition significantly right before merger \cite[e.g.,][]{Kanakis-Pegios:2022abm}. 
In addition, the crustal strain exerted by the mode reads $\sigma^2= 2(\boldsymbol{\sigma}_{ij})(\overline{\boldsymbol{\sigma}^{ij}})$, where overbar denote complex conjugation, and the stress tensor is given by \cite[see Eqs.~(22)-(23) in][]{Kuan:2021sin}
\begin{align}
    \boldsymbol{\sigma}_{\mu\nu}= \tfrac{1}{2}\left( \partial_{\mu} \xi_{\nu} + \partial_{\nu} \xi_{\mu} + \delta g_{\mu\nu} \right) - \Gamma^{\gamma}{}_{\mu\nu}\xi_{\gamma},
\end{align} 
with $\boldsymbol{\Gamma}$ the Christoffel symbols of the background metric, whose perturbation is $\delta \boldsymbol{g}$.
A given mode stresses the crust at a strength that is linear in the mode amplitude to which the crust may yield if a certain threshold is met. The threshold depends on the composition, including possible impurities, of the crust, as well as EOS; a range of maxima have been deduced from numerical simulations, varying from $0.04$ to $\sim0.1$ \citep{horo09,baiko18}. Here we adopt $\sigma_\text{max}=0.04$, i.e., crustal regions where a strain of $\sigma\ge\sigma_\text{max}$ are set to yield. We also work with the specific von Mises breaking criterion.

During and after shattering, some energy stored in the fractured crevice(s) is transferred to nearby regions, triggering aftershocks \citep{dunc98}, and to the exterior magnetic field. In the latter case, the energy deposited into  (open) field lines may lead to the transient gamma-ray emissions that constitute precursor flares. These emissions are expected to have non-thermal spectra if the field strength is sufficiently strong, $B \gg 10^{13}$ G \citep{Tsang:2011ad}. In the event that the precursor is accompanied by noticeable aftershock-induced mode(s), the light curve of the precursor may feature a quasi-periodic behaviour \citep{suvo22}, such as was observed in the recent event GRB 211211A \citep{gao22}. This latter event, despite being of long duration, was also accompanied by a kilonova and thus likely resulted from a merger event. We remark therefore that the method presented here may also apply to some long GRBs with double precursors, such as GRB 190114C (in principle), for which precursors were observed 5.6s and 2.9s prior to the main event \citep{copp20}.

\subsection{Binary formation channels} \label{sec:form}
Although the criterion {we adopt} for a crust yielding is simply that $\sigma_{\rm max}$ is exceeded {somewhere}, a number of factors complicate the overall resonant-shattering picture. Most notable for our discussion are issues related to the orientations, masses, and the EOS of both stars.
These factors hint at which star is more likely to generate precursors; for example, $g$-modes in stars with mass $\sim1.45M_{\odot}$ couple weakly to the exterior tidal field \cite[see Appendix of][]{prep}, and are thus incapable of producing resonant shattering flares. In addition, the misalignment angle will in general reduce the extent to which a mode can be excited. Therefore, we may be able to deduce which star in the binary exhibits precursors if these parameters are given. 

In the canonical formation channel of close binary NSs, there are two supernovae, likely of the \emph{ultra-stripped} variety, separated by a timescale that is sensitive to a number of source specifics \citep{tauris15,tauris17,Heuvel17}. Accretion by the first born from the not-yet-collapsed one can be of a disc or a wind-fed nature depending on the orbital period and Roche lobe interplay \citep{pac71}. If a disc forms, fluid motions in the magnetically-threaded disc torque the NS \cite[e.g.,][]{gs21}, gradually aligning its rotation axis with the angular momentum of the whole system while spinning it up. Wind-fed accretion is much less efficient for this purpose. The extent of alignment is therefore determined by the time separation between the two supernovae, and the nature of accretion (or its absence). The second NS will, however, not be `recycled' or have its angular momentum axis oriented with that of the orbit; even relatively weak supernova kicks may arbitrarily reorient the relevant axes \cite[cf.][]{tauris17}. Moreover, the coalescence will be expedited significantly if the companion is propelled towards the primary through a strong kick \cite[see][for a detailed discussion on the influence of natal kicks on the merger rate]{Postnov14}.
In some cases, the first born NS, taken as the primary in this article, can be long-term recycled and eventually reach an aligned, possibly large spin, while the companion is relatively slow with a potentially non-negligible tilt angle \cite[see, e.g.,][]{zhu18,zhu20}.
However, if accretion is prematurely truncated by the supernova of the companion or if both stars explode roughly at the same time, one may anticipate that the binary will consist of two misaligned NSs. If the coalescence time-scale turns out to be shorter than many spin-down times, the latter of which is likely controlled by magnetic braking, a system where the companion does not spin down significantly prior to coalescence could eventuate. 

To summarise, at moments close to merger, the primary may {rotate (comparatively) rapidly} and be aligned if it has undergone long-term recycling by disc-fed accretion, while the companion may be {rotating} relatively rapidly only if it is kicked towards the primary. Although the magnetic energy of the primary may be depleted by (Hall-accelerated) Ohmic decay \cite[e.g.,][]{smg16}, which brakes the star while it recycles, it may still have a strong, localised field `buried' under the accreted layers near the surface \cite[e.g.,][]{suvm20}. On the other hand, globally-strong fields could persist in the companion if it is kicked into the primary, especially if it settles into a `Hall attractor' state \citep{gour14}.
We note, additionally, that in the event of dynamical capture, both members of the binary are also likely to be misaligned, and the companion will be regardless. 

We stress that the above discussion is not meant to be an exhaustive survey on formation channel possibilities, but illustrates that a wide range of magnetic field strengths, spins, and misalignment angles are theoretically plausible. Observationally speaking, the tilt angle between the spin and the orbital angular momentum axes has been estimated for five Galactic NSs through geodetic precession and optical polarimetric measurements: $(18\pm6)^{\circ}$ for PSR B1913+16 \citep{kram98}, $<3.2^{\circ}$ for PSR J0737-3039A \citep{ferd13}, $(27\pm3)^{\circ}$ for PSR B1534+12 \citep{fons14}, $<34^{\circ}$ for PSR J1756-2251 \citep{ferd14}, and $>20^{\circ}$ Her X-1 \citep{doros22}.
Four have a mild misalignment angle.
Tidal activity in tilted NSs is investigated in Sec.~\ref{Sec:tilted}, using these data as representative examples.

\subsection{Resonance in Misaligned binaries}\label{Sec:tilted}

For the reasons discussed above, it is worth investigating how the resonant-shattering procedure proceeds in cases where one or both of the binary members are misaligned. Although the QNM excitation in an aligned NS is dominated by $m=2$ modes \citep{zahn77}, for a misaligned NS, in general, the whole range of $-l\le m\le l$ modes will be excited to an extent that depends on the Wigner $D$-functions defined below \cite[e.g.,][]{xu17}.
Without loss of generality, we focus on a tilted primary forced by the tidal potential built by the companion. The leading-order component of potential corresponds to $l=2$, and has the form
\begin{align}\label{eq:tidal_Phi}
    \Phi(\boldsymbol{x},t) =& -\frac{M_{c}r^{2}}{a^{3}}\left(\frac{3\pi}{10}\right)^{1/2}\nonumber\\
    &\times\left[e^{-2i\phi_c}Y_{22}(\theta_L,\phi_L)+e^{2i\phi_c}Y_{2,-2}(\theta_L,\phi_L)\right],
\end{align}
where $Y_{lm}$ denotes spherical harmonics, the numerical coefficient $\left(3\pi/10\right)^{1/2}$ is the constant $W_{2,\pm2}$ used in other studies [e.g., Eq.~(15) of \cite{kokk95}], and $\boldsymbol{x}$ denotes the spatial coordinates in the inertial-frame of the primary respect to which the phase coordinate of the companion is $\phi_c$. 
Here $(\theta_L,\phi_L)$ are the polar, and phase coordinates of the orbital angular momentum relative {to the spin-axis of the primary}. 

\begin{figure}
	\centering
	\includegraphics[width=\columnwidth]{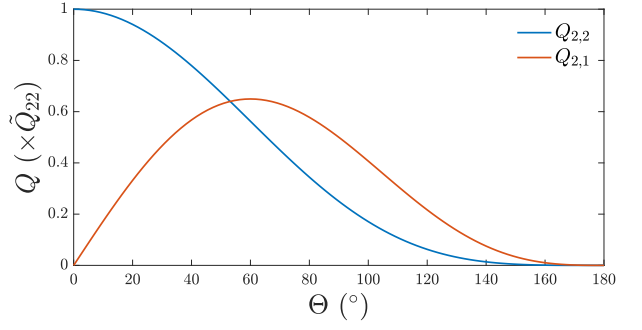}
	\caption{Tidal overlap of $m=1$ (red curve) and $m=2$ (blue curve) modes as functions of inclination, $\Theta$, in units of the tidal overlap of the $l=2=m$ mode in an aligned NS [i.e., $\tilde{Q}_{22}$; Eq.~\eqref{eq:Q22Q21}].}	\label{fig:wig}
\end{figure} 
\begin{table*}
	\centering
	\caption{Minimum tilt angle of the heavier NS, in binaries with a total mass of $2.5M_{\odot}$, such that the $m=1$ $g_1$-mode is excited to a greater extent than the $m=2$ $g_1$-mode in the sense that made clear in the maintext. Three mass ratios are considered, namely $q=1,0.9$, and $0.8$. For each mass ratio, we consider four spins ranging from 0 to 60 Hz.
	}
	\begin{tabular}{ccccccc}
	    \hline
	    \hline
		$\nu_{\star}$ (Hz), q && KDE0V & APR4 & SLy & ENG & MPA1 \\
		\hline
		(0,1--0.8) && $53.13^{\circ}$ & $53.13^{\circ}$ & $53.13^{\circ}$ & $53.13^{\circ}$ &$53.13^{\circ}$  \\
		\hline
		(20,1) && $59.15^{\circ}$ & $59.69^{\circ}$ & $59.83^{\circ}$ & $60.70^{\circ}$ & $61.90^{\circ}$ \\
		(20,0.9) && $59.23^{\circ}$ & $59.76^{\circ}$ & $59.76^{\circ}$ & $60.65^{\circ}$ & $61.30^{\circ}$  \\
		(20,0.8) && $59.01^{\circ}$ & $59.37^{\circ}$ & $59.40^{\circ}$ & $60.39^{\circ}$ & $61.39^{\circ}$ \\
		\hline
		(40,1) && $69.93^{\circ}$ & $72.01^{\circ}$ & $71.95^{\circ}$ & $76.03^{\circ}$ & $80.23^{\circ}$ \\
		(40,0.9) && $69.46^{\circ}$ & $71.54^{\circ}$ & $71.48^{\circ}$ & $75.18^{\circ}$ & $80.24^{\circ}$ \\
		(40,0.8) && $68.90^{\circ}$ & $70.97^{\circ}$ & $70.63^{\circ}$ & $74.55^{\circ}$ & $79.17^{\circ}$ \\
		\hline
		(60,1) && $92.68^{\circ}$ & $100.20^{\circ}$ & $100.64^{\circ}$ & $120.92^{\circ}$ & $146.87^{\circ}$ \\
		(60,0.9) && $90.65^{\circ}$ & $98.34^{\circ}$ & $97.79^{\circ}$ & $116.21^{\circ}$ & $142.90^{\circ}$ \\
		(60,0.8) && $88.90^{\circ}$ & $96.46^{\circ}$ & $95.52^{\circ}$ & $112.29^{\circ}$ & $139.35^{\circ}$ \\
	    \hline
	    \hline
	\end{tabular}
	\label{tab:1gt2}
\end{table*}

In terms of the Wigner $D$-functions, $D_{mm'}(\Theta)$, for tilt angle $\Theta$ made between spin and angular momentum axes, the potential \eqref{eq:tidal_Phi} can be rewritten as \citep{Lai06,xu17}
\begin{align}
    \Phi(\boldsymbol{x},t) =& -\frac{M_{c}r^{2}}{a^{3}}\left(\frac{3\pi}{10}\right)^{1/2}
    \bigg[e^{-2i\phi_c}\sum_{m'} D_{2m'}(\Theta)Y_{2m'}(\theta,\phi)\nonumber\\
    &+e^{2i\phi_c}\sum_{m''} D_{-2m''}(\Theta)Y_{2m''}(\theta,\phi)\bigg],
\end{align}
where the relevant $D$-functions read,
\begin{align}
    &D_{2,2}=D_{-2,-2}=\cos^4(\Theta/2), \\
    &D_{2,1}=-D_{-2,-1}=-2\cos^3(\Theta/2)\sin(\Theta/2)\\
    &D_{2,0}=D_{-2,0}=\sqrt{6}\cos^2(\Theta/2)\sin^2(\Theta/2)\\
    &D_{2,-1}=-D_{-2,1}=-2\cos(\Theta/2)\sin^3(\Theta/2)\\
    &D_{2,-2}=D_{-2,2}=\sin^4(\Theta/2).
\end{align}
We see that, in contrast to the non-spinning and aligned-spinning cases, modes with $m\ne2$ will also be excited by the tidal field in misaligned NSs to different levels depending on the inclination.

It is expected that the excitation of modes with $m\le0$ is weak relative to those with positive $m$ since spin reduces the frequencies of the retrograde modes, giving rise to earlier excitation or even resonance, where $\Phi$ is weaker. We therefore specify ourselves to modes with $m=1$ and $m=2$, for which the tidal overlap is, respectively, \citep{Kuan:2021jmk,miao23}
\begin{align}
    &Q_{22} = \frac{D_{2,2}}{M_{\star}R_{\star}^2}\int\sqrt{-g}d^3x (\rho+p)\xi^{\mu}e^{-2i\phi} \nabla_{\mu}(r^2Y_{22}),\\
    &Q_{21} = \frac{D_{2,1}}{M_{\star}R_{\star}^2}\int\sqrt{-g}d^3x (\rho+p)\xi^{\mu}e^{-i\phi} \nabla_{\mu}(r^2Y_{21}),
\end{align}
where the displacement $\xi^{\mu}$ is normalised such that
\begin{align}
	\int\sqrt{-g}d^3x (\rho+p)e^{-2\Phi}\xi^{\mu} \bar{\xi}^{\mu}=M_{\star}R_{\star}^2.
\end{align}
We note that the fluid motion caused by a general $m$-mode is $\xi^{\mu}e^{-im\phi}$, and thus modes with the same $l$ share the same displacement. 
Denoting the tidal overlap of an $l=2=m$ mode in an \emph{aligned} NS, where $D_{2,2}=1$ and $D_{2,1}=0$, as $\tilde{Q}_{22}$, we have the simple scalings 
\begin{align}\label{eq:Q22Q21}
	Q_{22} = D_{2,2}\tilde{Q}_{22}, \quad\text{and}\quad Q_{21} = |D_{2,1}|\tilde{Q}_{22}.
\end{align}
The relative strength between $Q_{22}$ and $Q_{21}$ is thus encoded in the associated Wigner $D$-functions, and depends on $\Theta$.

In Fig.~\ref{fig:wig}, we plot $D_{2,2}$ and $|D_{2,1}|$ as functions of $\Theta$, where we see that the overlap of the $m=1$ mode starts to be stronger than the $m=2$ one for $\Theta\gtrsim53.13^{\circ}$. We note additionally that $Q_{21}$ is less than half of $Q_{22}$ even when $\Theta \lesssim 40^{\circ}$. Here we consider the stratification of $\delta=0.012$. However, we should keep in mind the stronger overlap of $m=1$ does not necessarily imply that the excitation of an $m=1$ mode would be stronger than the $m=2$ mode because the mode frequency of the latter is lower, implying an earlier excitation. 
The exact value of the minimal $\Theta$ such that the excitation of $m=1$ mode is stronger (i.e., the saturation amplitude is larger) than the $m=2$ one depends on the EOS, mass ratio, and the nature of modes (e.g., $f$-, $g$-, or $i$-modes).
Taking $g_1$-modes for example, for binaries with total mass of $2.5M_{\odot}$ we list in Tab.~\ref{tab:1gt2} the minimal $\Theta$ for which the $m=1$ excitation in the heavier NS can exceed the extent of that of the $m=2$ mode for several mass ratios between two stars $q\le 1$, some spins of the breathing star, and five EOS. The considered EOS are arranged in the order of softness from stiffest (left; KDE0V) to softest (right; MPA1).
We see that the critical tilt angle admitting a greater dominant $m=1$ is not overly sensitive to the mass ratio, though we note that the dependence is enhanced when a larger spin is considered. In addition, the critical angle is larger for softer EOS since the $g_1$-mode's frequency is lower for a fixed stellar mass thus, the onsets of $m=1$ and $m=2$ excitations will be further separated.

\begin{figure}
	\centering
	\includegraphics[width=\columnwidth]{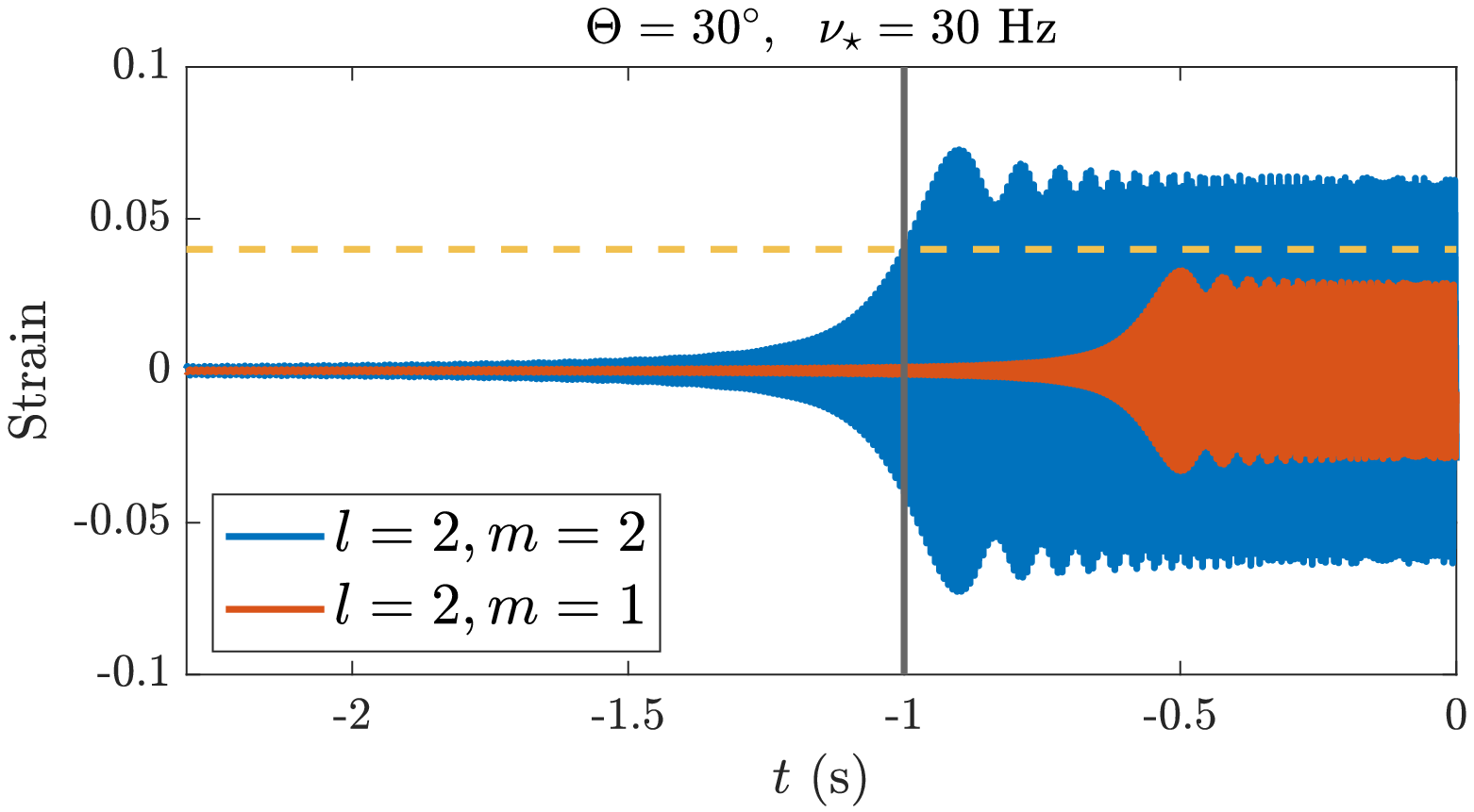}
	\includegraphics[width=\columnwidth]{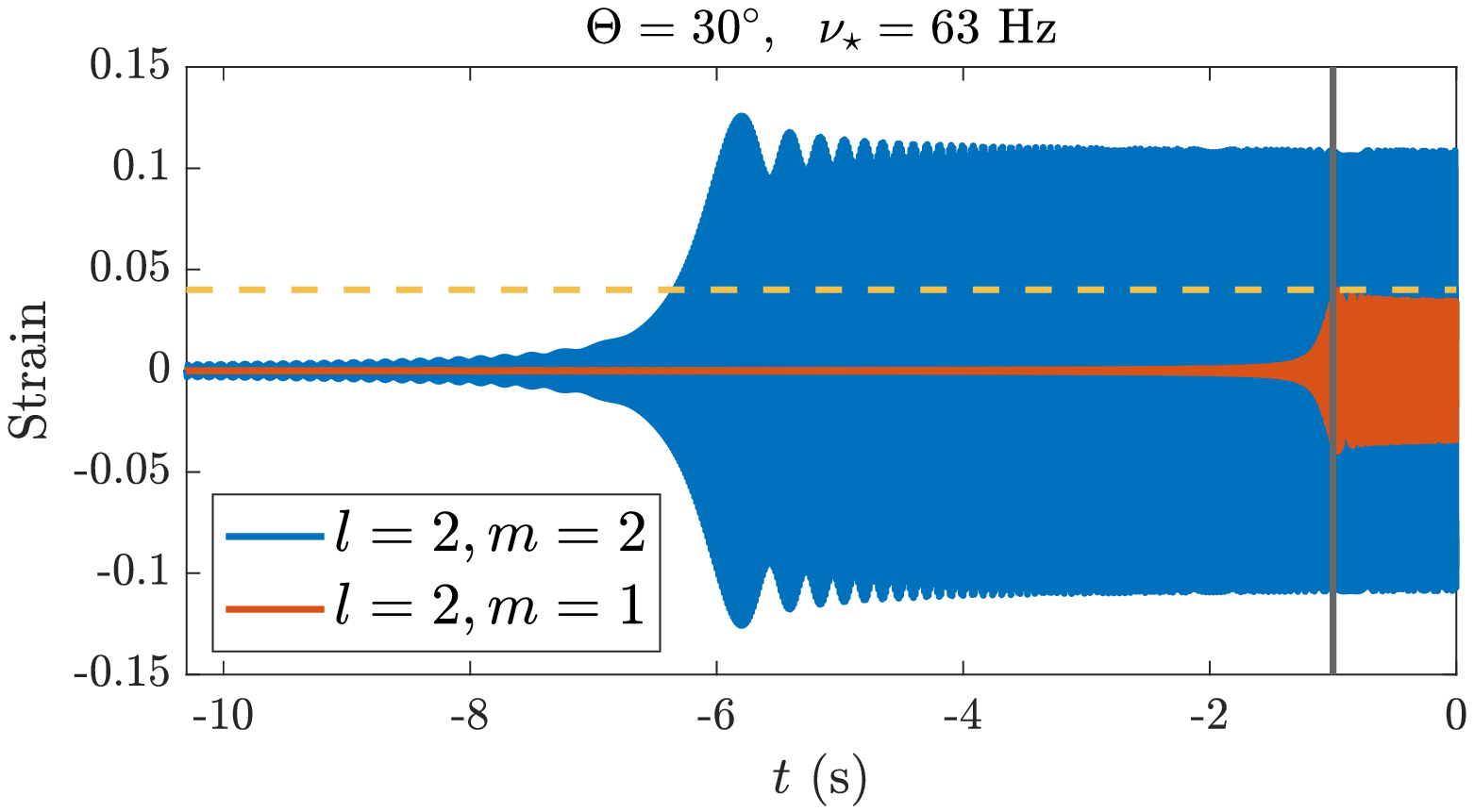}
	\includegraphics[width=\columnwidth]{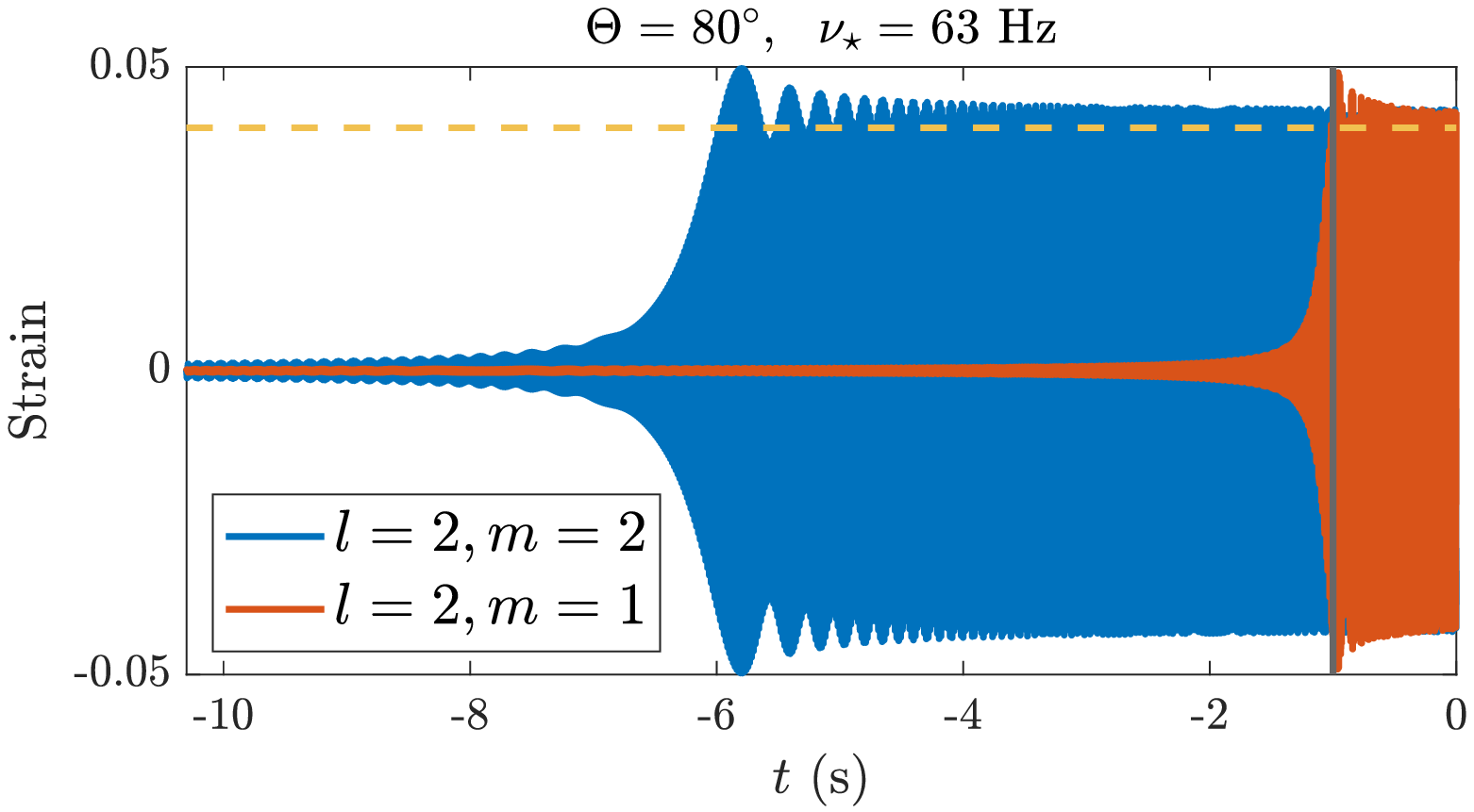}
	\caption{Strain ($\sigma$) induced by $m=2$ and $m=1$ $g_1$-modes (blue and red curves, respectively) of a star spinning at $\spin=30$~Hz (top panel) or at $\spin=63$~Hz (middle panel), both with $\Theta=30^{\circ}$, and a star spinning at $\spin=63$~Hz with $\Theta=80^{\circ}$ (bottom panel), all as functions of $t$, where the merger corresponds to $t=0$. The stratification is set through $\delta=0.021$, as relevant for the later precursor of GRB 090510 (see Sec.~\ref{case}). The horizontal dashed lines represent the breaking threshold; $\sigma_{\rm max} = 0.04$ is adopted here \citep{baiko18}. The binary is considered to be symmetric, and consists of stars of masses $1.23M_{\odot}$ with the APR4 EOS.
	}
	\label{fig:mis}
\end{figure}

We plot in Fig.~\ref{fig:mis} the evolutions of the strain by $m=1$ and $m=2$ $g_1$-modes in a NS member of an equal-mass binary as functions of the time prior to merger $t_p$. Three cases are shown: (i) The resonance of the $m=2$ mode occurs at $t_p=1$ s, whose strain at that moment equals $\sigma_\text{max}=0.04$ \citep{baiko18}. (ii) Same as (i) but for the $m=1$ mode. (iii) The resonance of the $m=1$ mode occurs at $t_p=1$ s, whose saturation strain equals that of the $m=2$ mode.
In each of these three scenarios, the strain of $m=2$ mode successfully exceeds the cracking threshold, while the $m=1$ mode is not strong enough to break the crust for case (i).
In the middle and bottom panels [case (ii) and (iii)], we see a $60\%$ drop in the saturation strain of the $m=2$ mode and a mild ($\sim 25\%$) increase in that of the $m=1$ mode by increasing the tilt angle from $30^{\circ}$ to $80^\circ$. 
From Fig.~\ref{fig:wig} we know that the maximal strain of both modes will decrease monotonically as $\Theta\sim 60^{\circ}$ for a given spin. Actually, for the star considered in Fig.~\ref{fig:mis}, neither mode may be able to yield the crust if $\Theta\gtrsim 80^{\circ}$.

\subsection{Tidal heating}\label{tidheat}

\begin{figure}
	\centering
	\includegraphics[width=\columnwidth]{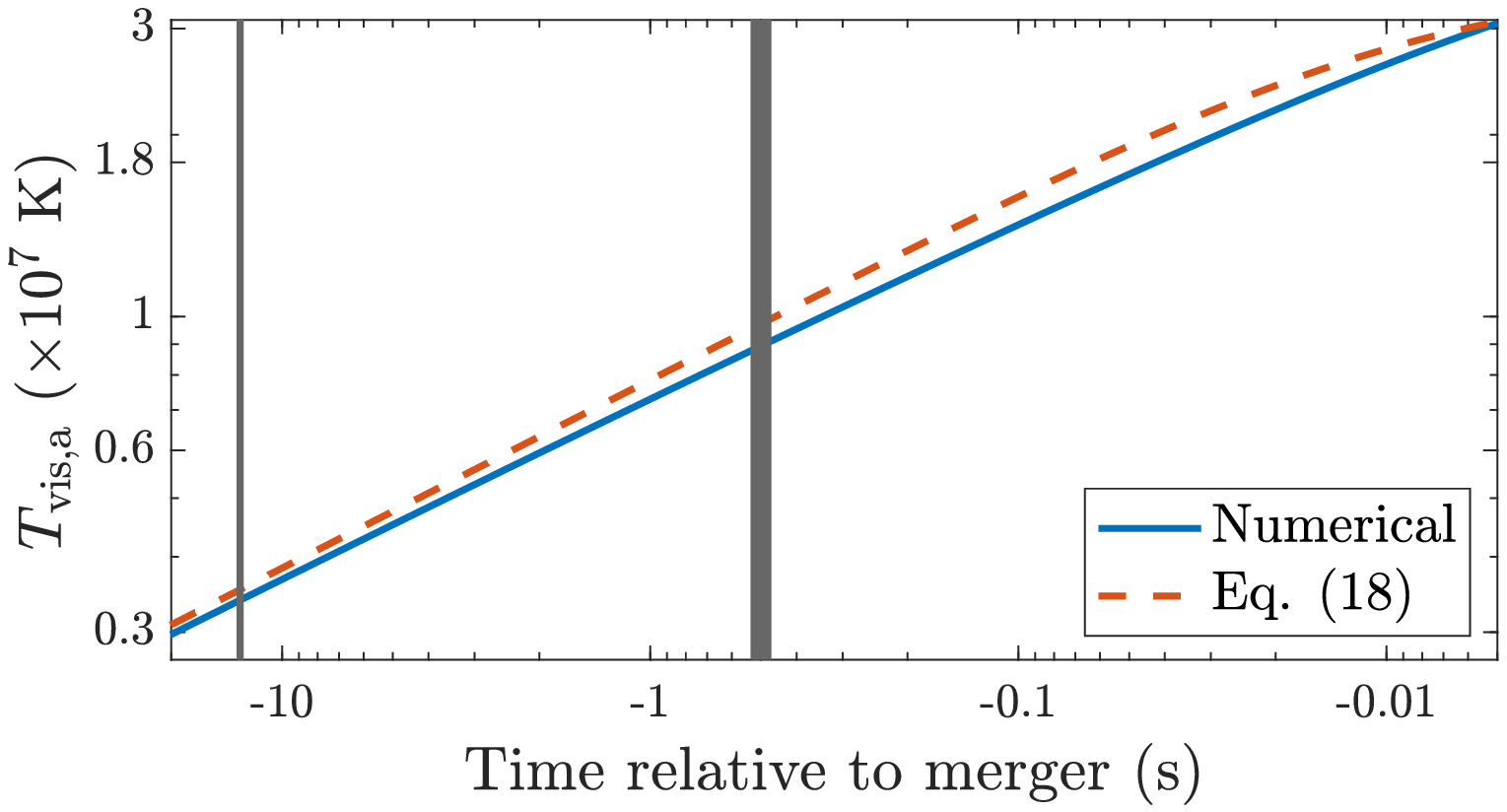}
	\includegraphics[width=\columnwidth]{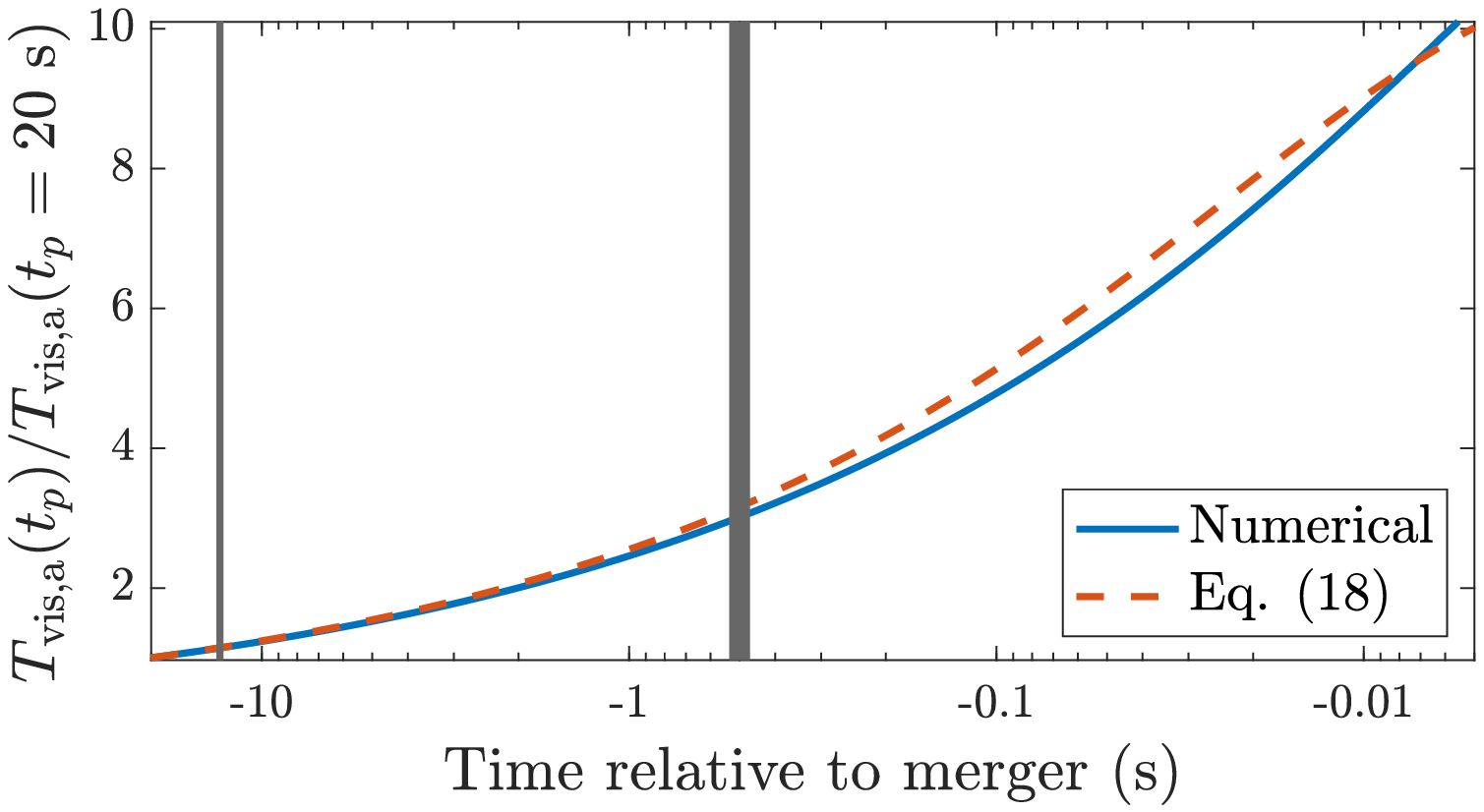}
	\caption{Top panel: Numerical (blue) and analytic [red; Eq.~\eqref{eq:approxT}] temperature evolutions due to heating by the viscous dissipation of $f$-mode excitations for a symmetric binary consisting of stars with $1.23M_{\odot}$ and EOS APR4. Bottom: The ratio of the temperature between a varying $t_P$ and $t_p=-20$~s corresponding to thermal ramping. 
	In both panels, the vertical lines mark the timing of the first and the second precursor of SGRB 090510, and the thickness of them notes the error of 50 ms for each timing.}
	\label{fig:heat}
\end{figure}

In addition to the alignment concern about the NS, tidal heating due to the viscous dissipation activated by mode excitations also complicates the present investigation. As described previously, the stratification is a function of time, $\delta(t)$, because of heating implying that the $g$-mode spectrum itself is time-dependent. 
In particular, \cite{lai94} has shown that a non-resonantly excited $l=2=m$ $f$-mode in an aligned NS will increase the star's temperature by [see Eq.~(8.30) therein]
\begin{align}\label{eq:heat}
    T_{\text{vis,a}} \approx 3.6\times10^{7}\left( \frac{3R_{\star}}{a} \right)^{5/4} \text{K},
\end{align}
which depends on the stellar radius and separation $a$; we remark that the subscript $a$ here denotes the heating in \textit{aligned} stars. 

{In the last stages of merger, $f$-modes lead to the greatest degree of heating \cite[$\sim5$ times more than $g$-modes, for instance; cf.][]{prep} because they efficiently vibrate the entire star. While $g$-modes actually lead to equal or even dominant heating over the life of the binary, their contribution is negligible relative to the $f$-mode in the last $\lesssim 10^{2}$s of inspiral where our simulation applies. However, this does not necessarily imply that $f$-mode heating controls the stratification index in regions of the star where the $g$-mode eigenfunction is defined (e.g., in the crust). Even if other modes carry less overall energy, for example, they could heat the crust to a larger degree than the global $f$-modes. This is especially true in cases where a $g$-mode breaks the crust, which goes on to experience plastic heating \cite[e.g.,][]{link96,belo16}. It is, in general, a difficult problem to assess and evolve the local temperature gradients that result from mode-induced perturbations, resonant or otherwise \cite[though see][]{pan20}. To provide a concrete but simple example, we adopt the volume-averaged expression \eqref{eq:heat} to define how the temperature, and hence $\delta$, evolves following the first fracture. A further study considering a more sophisticated temperature profile in the crust \citep[e.g.,][]{gudm82,link96,pote97} is deferred to future work.}

For a particular binary, we show $T_{\text{vis,a}}$ as a function of time in Fig.~\ref{fig:heat}. Although the $f$-mode excitation also modifies the orbit evolution, to estimate the heating to leading order we adopt the well-known expression for the separation of binaries shrinking due to quadrupolar gravitational emission generated by the orbital motion (mode backreaction is not included), given by
\begin{align}
	a(t_p) = \left(81R_{\star}^4+\frac{256}{5}M_{\star}^3q(1+q)t_p\right)^{1/4}.
\end{align}  
Substituting this into Eq.~\eqref{eq:heat} returns
\begin{align}\label{eq:approxT}
    T_{\text{vis,a}} \approx 3.6\times10^{7}\left[ 1+\frac{256M_{\star}^3q(1+q)t_\text{p}}{405R_{\star}^4} \right]^{-5/16} \text{K}.
\end{align}
Variations in $\delta$ obey the relation $\Delta\delta/\delta=2\Delta T/T$ since $\delta\propto T^2$, where $\Delta T\simeq T_\text{vis,a}$. {We stress again that this $\Delta T$ provides only a crude approximation to the degree of \emph{crustal} heat deposited between precursors, as it ignores localised heating due to $g$-mode excitations and the fracture itself \cite[see][for discussion]{lai94}.}

In this article, the primary is defined as the first formed NS in a binary instead of the heavier one, while the mass ratio is defined as the ratio between the mass of the lighter NS and that of the other. 
In principle, there is a non-linear influence of evolving $\delta$: $g$-modes become resonant at different times, thus the orbital frequency decays differently. Here we ignore this latter non-linear effect (cf. $p$-$g$ mode couplings also; see Sec. \ref{grawave}).

In the top panel of Fig.~\ref{fig:heat}, we plot the heating \eqref{eq:heat} using a numerically solved separation $a(t)$ \citep[see, e.g.,][for numerical details on the standard Hamiltonian treatment]{Kuan:2021jmk} together with the analytic expression \eqref{eq:approxT} for a particular binary.
We see that the analytic form matches to the numerical results to within $5\%$ until the final 5 s, where the difference gradually reaches $10\%$ over the next $4.9$ s until rapidly growing to $20\%$ in the last 0.1 s.
The consistency between numerical and analytical results reinforces the applicability of the analytic formula for our purposes, especially since we care only about the ratio of viscosity-generated increases in temperature, $T_\text{vis,a}$, between different times. In the bottom panel of Fig.~\ref{fig:heat}, we plot the ratio of $T_\text{vis,a}$ between a varying time and the moment $t_p=20$ s. 

The above equations apply to aligned NSs, while for misaligned system, the tidal heating will be modified by the inclination $\Theta$. In the adiabatic limit adopted in \cite{lai94}, the $m\ne\pm2$ $f$-modes heat up the star in the same manner, leading to the $\Theta$-modulated expression:
\begin{align}\label{eq:vis_misaligned}
    T_{\text{vis}} \approx T_{\text{vis,a}}\sum_{m=-2}^{2}D_{2,m}(\Theta),
\end{align}
where we see a reduction in the tidal heating since $\sum_{m=-2}^{2}D_{2,m}(\Theta)<1$ unless the star is aligned or totally misaligned. At $\Theta=90^{\circ}$, where the reduction in the heating is the strongest, the increase in temperature is only $\gtrsim11\%$ relative to the $\Theta=0$ or $180^{\circ}$ cases. Despite the modified temperature evolution, the ratio between $T_\text{vis}(t)$ and $T_\text{vis}(t=-20)$ is the same as aligned systems. The weakened $f$-mode excitation may also change the inspiral trajectory, and thus the time before merger $t_p$ cannot be trivially compared from case to case. However, this effect results in at most $5$ radians of dephasing in the gravitational waveform for $\spin\lesssim100$ Hz \citep{kuan22}, causing a $\ll 1$ s error in the merger time prediction, so we ignore such complications here.

\section{Spin frequency determination from precursor doubles}
\label{Sec:spinfreqdet}

In this article, we assume that both precursors are set off from one star -- either the primary (the one that forms earlier; see Sec.~\ref{shattering}) or the companion -- and further that they are attributable to $g_{1}$ and $g_{2}$ resonances. There are other theoretical possibilities, however, notably that non-$g$ modes are responsible or that each star releases a flare at different times rather than one star releasing both. These are discussed in detail in Sec.~\ref{sec:diff}.

The orbital frequencies at which two precursors A and B are observed, denoted by $\nu_A$ and $\nu_B$ with $\nu_A>\nu_B$, should be determined by their preceding time relative to the merger, while the measured quantity is the waiting time, i.e., the preceding time relative to the main burst. Therefore, $\nu_A$ and $\nu_B$ depend on the unknown jet formation timescale $\JT$ for the associated SGRB when the waiting time are given. In addition, a resonant overstraining will not instantaneously lead to a precursor and there is some time delay between $g$-mode resonance and flare. In particular, there are two times to consider in a failure-induced flare scenario: (i) the time taken for the crust to actually fail following an overstraining from resonance, and (ii) the time for emissions to be generated following a failure (see also \citealt{thom95} for a discussion in the magnetar flare context). \cite{Tsang:2011ad} estimate (i) to be $\sim 1$~ms based on elastic-to-tidal energy ratios, though the value depends on both the overlap integral and mode frequency (see below Eq.~10 therein). For the latter, \cite{neil22} argue that the timescale could be as long as $\sim 0.1$~s for $B \sim 10^{13}$~G; see their Eq.~(12). However, this estimate assumes $\sigma_{\rm max} \sim 0.1$ estimated from \cite{horo09}, though a more recent study by \cite{baiko18} finds $\sigma_{\rm max} \approx 0.04$. We estimate, following \cite{neil22},
\begin{equation}
t_{\rm emit} \sim \frac{E_{\rm elastic}}{L_{\rm max}} \approx 0.03 \left(\sigma_{\rm max}/0.04\right)^2 (L_{\rm max} / 10^{47} \text{erg/s})^{-1} \text{ s},
\end{equation}
where $L_{\rm max}$ is the rate at which energy can be extracted by the magnetic field. For even modestly bright precursors [cf. the precursor in GRB 211211A, with luminosity reaching $\sim 7 \times 10^{49}$ erg/s \citep{xiao22}], $t_{\rm emit}$ is therefore sub-leading with respect to the observational uncertainties already present in estimating the precursor onset time \citep{copp20,Wang:2020vvr}.
(Note that we also ignore temperature changes to the lattice strain threshold, which reduces $\sigma_{\rm max}$, and thus $t_{\rm emit}$ is an overestimate for late-time precursors). In Table 2 below and throughout, we account for a $\gtrsim 0.1\,$s tolerance in the onset time, absorbing uncertainties related to the two timescales described above.

In this work, the orbital evolution is numerically simulated by using a 3rd-order post-Newtonian (PN) Hamiltonian, a 2.5 PN treatment for GW back-reaction, and the tidal effects of the $l=2=m$ $f$-modes \cite[see][for details]{Kuan:2021sin}.
Following this notation, we denote the stratification indices at the time of the precursors as $\delta_A$ and $\delta_B$.
Matching the tidal-driving frequency to the inertial-frame frequencies of $g_1$- and $g_2$-modes returns [from Eqs.~\eqref{eq:g1g2} and \eqref{eq:rotshift}]
\begin{align}\label{eq:spin}
    \spin(\nu_{A,B},\delta_{A,B},\JT) \approx \frac{1.24\nu_A(\JT)-2\nu_B(\JT)\sqrt{\delta_A/\delta_B}+\beta}{m_2(1-C_2)\sqrt{\delta_A/\delta_B}-0.62m_1(1-C_1)},
\end{align}
where $m_{1}$ and $m_2$ are the winding numbers associated with the $g_{1}-$ and $g_{2}-$mode that accounts for the precursor, and the numerical coefficients come from the fitting parameter in Eq.~\eqref{eq:g1g2} and the respective integrated constants $C$ defined in \eqref{eq:rotshift}.

Although relation \eqref{eq:spin} holds for the EOS in Fig.~\ref{fig:g1g2}, we, hereafter, specify ourselves to those EOS able to support stars with mass $\gtrsim2M_{\odot}$ or more so as to be consistent with Shapiro delay measurements of PSR J0740+6620 \citep{crom20,fons21}. These are listed in Fig.~\ref{fig:spn}. 
This requirement for the maximal mass attainable of a certain EOS is conservative since we take the mass of this millisecond pulsar as a potential limit of static NSs, while J0740 spins at $\sim346$ Hz \cite[see Tab.~1 of ][]{crom20}. It is worth pointing out the recent discovery of PSR J0952-0607 may set a novel record on mass of NSs at $M=2.19M_{\odot}$, which has a higher spin of $\sim700$ Hz nonetheless. 
In addition, we note that relation \eqref{eq:spin} does \emph{not} depend on $\Theta$, and acts as a \emph{necessary condition} for double precursors but not a sufficient condition. The other conditions required within the resonance shattering scenario of $g$-modes are: 
(i) the excitation occurring at the precursor timing should be strong enough to yield the crust, thus requiring a minimum magnitude to the tidal overlap\footnote{It has been shown in the Appendix of \cite{prep} that the $g$-modes of stars with mass close to $1.45M_\odot$ are less susceptible to the tidal field, and thus are less likely to break the crust.} and maximum on the inclination $\Theta$,  and 
(ii) the energy stored in the cracking area should be large enough to accommodate the luminosity of the precursor(s).

We here do not take into account $m\le 0$ modes as discussed in Sec.~\ref{Sec:tilted}; accordingly, the axial quantum numbers of the modes for both precursors may take values of $1$ and $2$, leading to four possible combinations. As far as the inferred spin is concerned, the two extreme cases are, respectively, $(m_1,m_2)=(1,2)$ (smallest estimate) and $(2,1)$ (largest estimate). Although the difference can be larger than $100\%$ (cf.~Tab.~\ref{tab:m1m2}), narrowing down the choice to only one or two can be, in principle, achieved when combining these data with other observations coming from, e.g., the kilonova \citep{papan22}, the presence or absence of spin-induced phase shifts in gravitational waveform \citep{stein21,kuan22}, and from the nature of the accretion disk surrounding the merger site \citep{east19}. 
In addition, the free mode frequencies obtained retroactively from the spin estimate should be accessible (see below).

\subsection{Case study: GRB 090510}\label{case}
\begin{table}
	\centering
	\caption{Spin predictions, in Hz, for a variety of different resonance scenarios (i.e., for different azimuth number combinations; last four columns) and relative precursors timings ($t_p$; first column) in a GRB095010-like system with a tilted binary (i.e., $\Theta \ne 0^{\circ}$; though note that the angle does not affect the inertial-frame mode frequency). We take the EOS as APR4 and fix the primary mass to $1.4 M_{\odot}$, assuming an equal-mass companion $(q = 1)$.
}
	\begin{tabular}{cccccc}
	    \hline
	    \hline
	    Relative precursor timings && \multicolumn{4}{c}{($m_1,m_2$)} \\
		$t_p$ (s) && (1,1) & (2,1) & (1,2) & (2,2) \\
		\hline
		0.5 , 13 && 6.55 & 13.11 & 2.62 & 3.27 \\
		0.45 , 13 && 9.54 & 19.11 & 3.82 & 4.77 \\
		0.55 , 13 && 3.92 & 7.84 & 1.57 & 1.96   \\
		0.5 , 12.5 && 5.41 & 10.83 & 2.16 & 2.71 \\
		0.5 , 13.5 && 7.62 & 15.25 & 3.05 & 3.81 \\
	    \hline
	    \hline
	\end{tabular}
	\label{tab:m1m2}
\end{table}

GRB 090510 displayed two precursors at $\sim 13$ s and $\sim 0.5$ s prior to the main event \citep{abdo09,Troja:2010zm}. Neglecting the jet formation timescale for now, i.e., taking $\JT=0$ ms, we find $\nu_A = 68-80$~Hz and $\nu_B = 22-25.5$~Hz over a wide range of binaries for each of the considered EOS: those with total mass $M_{\text{tot}}=2.5-3.1M_{\odot}$ and mass ratios such that the lighter NS is heavier than $1M_{\odot}$ (see below). With the complication introduced by the axial number discussed above in mind, we present results assuming $m_1=2=m_2$ in this section, while discussion pertinent to other combinations is provided when appropriate.
The $m_{1}=m_{2}=2$ assumption is especially applicable to a star with an inclination $\Theta\lesssim 40^{\circ}$, while it becomes less feasible if the star is strongly tilted (cf.~Fig.~\ref{fig:wig}).

Assuming $\delta=0.005$ at 20 s prior to merger, which is controlled by the compositional stratification mostly and suitable for a mature NS before a potential tidal heating \citep{xu17}, the heating obtained via Eq.~\eqref{eq:approxT} gives the stratifications $\delta_{A,B}$ at the time of each precursor. An analysis over a uniform spread of stellar masses and radii, spanning $1-2.2M_{\odot}$ and $10.5-13$ km, and mass ratio over $0.7-1$, reveals that the inferred values of $\delta_{A,B}$ are rather insensitive to these three parameters, with the strongest dependence being on $M_{\star}$. Over the whole parameter space, we find $\delta_A\simeq0.021$ and $\delta_B\simeq0.006$ with errors of only $10^{-4}$ for $\delta_A$, and $10^{-7}$ for $\delta_B$. {We caution the reader that the narrow error bars on both $\delta_A$ and $\delta_B$ result from our simple heating model (Sec.~\ref{tidheat}), which may not approximate well the situation where crustal heating is treated rigorously as in, e.g., \cite{ripe91,pan20}.} 
The stratifications at the occurrences of the two precursors therefore do not depend on the aforementioned three quantities except in a mild way, implying that we can measure the spin of the NS hosting the double regardless of whether it is the primary or companion.

After fixing the respective values of $\delta$, the denominator of Eq.~\eqref{eq:spin} is found to be roughly constant: for the aforementioned prior of stellar parameters, the denominator\footnote{The mean for combinations $(m_1,m_2)=\{(1,2)$, $(2,1)$, $(1,1)\}$ is collectively listed as $\{2.7629,0.5519,1.1049\}$ with respective standard deviations of $\{0.0129,0.0068,0.0064\}$. These values can be substituted in the denominator of Eq.~\eqref{eq:spin} to get the formula similar to Eq.~\eqref{eq:surrogate} suitable for the associated combination of $(m_1,m_2)$.} has a normal distribution with a mean of 2.2099 and a standard deviation of 0.0128.
This fact simplifies Eq.~\eqref{eq:spin} to one solely in terms of the orbital frequencies as
\begin{align}\label{eq:surrogate}
	\spin(\nu_A,\nu_B,\JT) \approx 0.45\times[1.24\nu_A(\JT)-3.74\nu_B(\JT)+\beta].
\end{align}
The spin for the combination $m_1=2=m_2$ is, therefore, well approximated as a function of the chirp mass $\mathcal{M}$ alone, in the form
\begin{align} \label{eq:chirp}
    \spin \approx 3.28 \left(\mathcal{M}/1.19M_{\odot}\right)^{-0.44} \text{ Hz}.
\end{align}
The above holds because the orbital evolution is to a large measure determined by $\mathcal{M}$, even though tidal forces and the stellar $f$-modes influence orbital dynamics; this is also the main reason why, in GW analysis, $\mathcal{M}$ can be measured rather accurately. For a `canonical' case with $\spin=3.28$~Hz, the free mode frequencies of the $g_1$- and $g_2$-modes are found to be, respectively, $\sim145-165$~Hz for $\delta=\delta_A$ and $\sim50-57$~Hz for $\delta=\delta_B$ depending on $M_\star$ and EOS. 
An important aspect to note is that an MPA1 NS, having frequencies in the aforementioned ranges for the $g_1$- and $g_2$-modes, is close to the range of tidal-neutral models. In this sense, this EOS may be in tension with this scenario for the double precursors with timing similar to the two of GRB090510.

Although we focus on the combination of $m_1=2=m_2$ here, the variation in the inferred spin due to different combinations is explored in Tab.~\ref{tab:m1m2}, where we summarise the predicted spin of an APR4 NS having $1.4M_{\odot}$ in a binary with $q=1$ for GRB090510 under all possible combinations, including possible errors on the precursor timing. We see that for a given timing of the double precursors (i.e., the same row in Tab.~\ref{tab:m1m2}), the spin can have four possible values which vary by over $100\%$. In addition, an uncertainty of $0.1$~s in the timing of the later (i.e., smaller $t_{p}$) precursor can lead to an error of $\gtrsim10$~Hz in the spin estimate as shown in the first three rows of Tab.~\ref{tab:m1m2}. The reason behind this susceptibility is that the orbital frequency changes rapidly in the last stages before merger; in particular, an uncertainty of $\sim 0.1$~s between $t_p=0.45$ and $t_p=0.55$ corresponds to a increase of $\gtrsim5$~Hz in the orbital frequency (see Sec.~\ref{sec:jet} for more detail). Similar changes in the predicted spin are observed if the earlier precursor has a timing error of $\gtrsim 1s$.

\begin{figure}\label{fig:nu_q}
	\centering
	\includegraphics[width=\columnwidth]{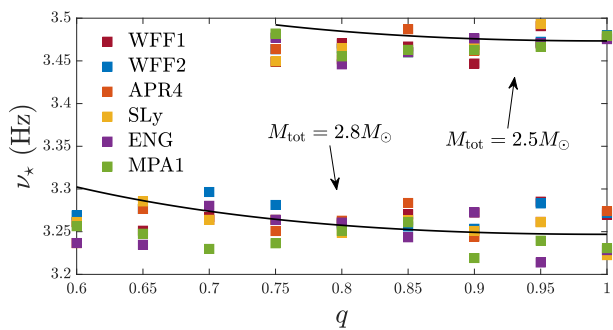}
	\caption{Derived spin of the primary [Eq.~\eqref{eq:surrogate}] for double-precursor events with various EOS in binaries with fixed total masses $2.5M_{\odot}$ and $2.8M_{\odot}$ as functions of mass ratio $q$. Solid lines represent equation \eqref{eq:chirp}.}
	\label{fig:spn}
\end{figure}

Taking two specific binary sequences, each characterised by a fixed total mass, we plot $\spin$ in Fig.~\ref{fig:spn} as a function of the mass ratio, $q$, where the solid lines represent the respective fittings \eqref{eq:chirp}. We see that $\spin$ depends only weakly on both $q$ (differs by less than 1 Hz between $q=1$ and $q=0.75$) and the EOS.
For $M_{\text{tot}}=2.5M_{\odot}$, one may expect the remnant to be supra-massive or even stable for the MPA1 EOS (stiffest one studied), surviving collapse long enough to produce an X-ray afterglow (see Sec. \ref{after}), as appropriate for GRB 090510. For this total mass, we do not consider values $q<0.75$ since $q\approx0.75$ implies a very light companion with $\sim1.07M_{\odot}$; this would be in tension with the lightest known NS, viz.~the secondary of J0453+1559 \cite[$1.18M_{\odot}$;][]{Martinez:2015mya}. By contrast, a hypermassive remnant may be expected for $M_{\text{tot}} \gtrsim 2.8 M_{\odot}$.
It is worth mentioning that the lighter NS for the considered supra-massive remnant cases is more susceptible to the tidal push since $g$-modes in NSs with mass closer to $1.45M_{\odot}$ are more tidally neutral \cite[see Appendix A of][]{prep}.

Although the inference of spin is largely insensitive to $q$, the extent to which a mode can be excited depends on $q$. Together with $\Theta$, $M_{\star}$, $q$, and EOS, these factors are core to the rather involved problem of whether the generated strain is strong enough \cite[which also depends on the crust model; see][]{baiko18}. Since the present article aims to point out the necessary condition \eqref{eq:spin}, we defer this complicated, muli-dimensional issue to a future study.

\subsection{Jet delay corrections}
\label{sec:jet}
\begin{figure}
	\centering
	\includegraphics[width=\columnwidth]{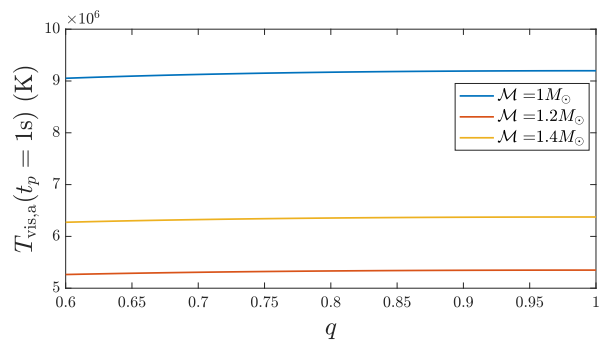}
	\caption{Compactness-scaled $T_\text{vis,a}$, as a function of $q$, for three representative chirp masses, ${\cal M}$. EOS MPA1 is used to obtain $R_\star$ from the stellar mass $M_\star$ derived from ${\cal M}$ and $q$. }
	\label{fig:Tq}
\end{figure}

Allowing for a non-zero jet formation and breakout timescale, $\JT$, gives rise to a shift $\delta\spin$ in the spin inference from equation \eqref{eq:chirp}. 
Since the binary evolution is largely determined by $\mathcal{M}$, it is expected that the mode frequencies associated with a given precursor timing is a function of $\mathcal{M}$ to leading order. That said, $\nu_A$ and $\nu_B$ in Eq.~\eqref{eq:spin} can be estimated accurately if ${\cal M}$ is given, while the stratifications relevant to the tidal heating are subject to the influence of the mass ratio, as suggested in Eq.~\eqref{eq:approxT}. 
However, for a fixed ${\cal M}$, the dependence of $T_\text{vis,a}$ on $q$ is only slight. In particular, the dependence on $q$ is encoded in the combination $M_\star^3q(1+q)/R_\star^4$, which is rather insensitive to $q$ as shown in Fig.~\ref{fig:Tq}. While in Fig.~\ref{fig:Tq} we only show the results for EOS MPA1, we note that the dependence on $q$ is weak for all the considered EOS.
For example, the heating difference in the heavier (or equally heavy) star between $q=1$ and $q=0.7$ cases is only $\sim2.5\%$. The weak influence of $q$ on the ratio $\sqrt{\delta_A/\delta_B}$ is at a roughly the same level since $\delta\propto T^2$. In addition, this deviation is independent of ${\cal M}$. 
Accordingly, equation \eqref{eq:surrogate} is modified as
\begin{equation}\label{eq:surrogate_v2}
\begin{aligned}
	\spin(\nu_A,\nu_B,\JT) \approx& 0.45\times \Big[ 1.24\nu_A(\JT) \\
	&-(3.65\pm0.09)\nu_B(\JT)+\beta \Big]
	\end{aligned}
\end{equation}
to engulf the uncertainty of $q$ in the range of $0.7\le q\le1$.

For $\JT\lesssim 200$ ms \citep{Zhang:2019ioc}, and for the specific situation detailed in Sec.~\ref{case}, we find the relation weakly depends on $\mathcal{M}$:
\begin{equation}\label{eq:jet}
\hspace{-0.2cm}\begin{aligned}
&   \delta\spin/\spin \approx (0.83\pm0.13)\left(\frac{\JT}{100\text{ ms}}\right)
    +(0.20\pm0.03)\left(\frac{\JT}{100\text{ ms}}\right)^2 \\
     &-\left[(0.11\pm0.03)\left(\frac{\JT}{100\text{ ms}}\right)     +(0.03\pm0.01)\left(\frac{\JT}{100\text{ ms}}\right)^2\right]\frac{{\cal M}}{M_{\odot}},
\end{aligned}
\end{equation}
where the error budgets in coefficients are due to the uncertainty of precursor timing.
If the SGRB takes less than $20$ ms to launch and break out, for instance, the correction is $\lesssim15\%$ while the exact value depends on ${\cal M}$ and $q$. 

\section{Connections to other observational channels}
\label{sec:otherobs}

In this section, we explore some connections that the double precursor scenario has to GWs (Sec.~\ref{grawave}) and X-rays afterglow (Sec.~\ref{after}), and also provide some discussion about how non-$g$-mode scenarios can still be used to constrain the stellar properties (Sec.~\ref{sec:diff}).

\subsection{Gravitational waves}\label{grawave}

There are two phases for which GW measurements may augment our knowledge about systems with precursors: during the merger and from the remnant. During merger, tidal resonances, and forces more generally, accelerate the inspiral. These influences on the waveform may be connected back \ properties of the pre-merging stars to infer not only $\mathcal{M}$, $q$, and the effective spin parameter of binaries $\chi_\text{eff}$ \citep{zhu20} but also the stellar compactnesses via the mutual deformability, $\tilde{\Lambda}$. Generally speaking however, $\tilde{\Lambda}$ can only be tightly constrained by using priors for $\spin$ \citep{gw170817,ann18} since (i) the spins also induce certain dephasing in the gravitational waveform, degenerate with that caused by tidal activities, and (ii) the spin-modulated QNM spectrum may enhance the tidal contribution in the dephasing \citep{stein21,kuan22}. From the waveform (de)phasing, it is also possible to constrain the influence of unstable couplings between $p$- and $g$-modes \citep{lvk_pg,reye20}, where the amplitude is collectively set a collection of excited $p$-$g$ pairs \citep{wein13,Essick16}.

In addition, gravitational radiation from the remnant, which may be observed either directly or via the fall-off slope of electromagnetic emissions \cite[see Sec. \ref{after};][]{gl16}, allow us to infer properties of the final star. It was shown by \cite{man21} that many stellar parameters, such as the compactness, of a long-lived remnant NS (i.e., when $M_{\text{tot}} \lesssim 2.5 M_{\odot}$; see Fig. \ref{fig:spn}) can also be inferred from the mutual tidal deformability. Furthermore, the mass of the remnant may be reliably estimated from the chirp mass to within an error of at most $\lesssim 0.1M_{\odot}$ \citep{Bauswein:2015vxa}.  The frequency of the $f$-mode, from which independent constraints on the EOS can be placed, in the remnant can also be predicted if the spin frequency is known \citep{Kruger:2019zuz}. Additionally, large pre-merging spins may result in high degrees of mass asymmetry in the remnant \citep{papan22}, possibly revealing itself through the so-called ``one-arm'' instability in the GW spectrum or shifting the bar-mode peak \citep{east19}. Under favourable orientations, {(unstable)} QNMs from the {rapidly-spinning} remnant may be observable with the Einstein Telescope out to $\gtrsim 200$~Mpc \citep{don15}.

\subsection{Afterglow light-curves}\label{after}

GRB 090510 (and many other SGRBs) displayed an afterglow `plateau', which suggests that a NS was born following the merger \citep{ciolfi20}. Depending on the compactness and spin-down radiation efficiency of the remnant, analyses of the light curve indicate that the newborn star had a period in the range $1.8-8$ ms, surface magnetic field strength of $(5-17) \times 10^{15}$~G, and quadrupolar ellipticity between $10^{-4}$ and $10^{-2}$ \citep{rowl13,suvk21a,suvk21b}. These features impact the potential GW signal; for example, the characteristic strain $h_{0} \propto \epsilon \spin^2$. From a purely electromagnetic standpoint, an eventual falloff slope of $-2$ in the X-ray emissions would be expected for dipolar spindown, while GW-dominated energy losses would be characterised by a slope of $-1$ instead, the crossover time and luminosity of which can be used to infer the ellipticity and {(surface)} magnetic field strength \citep{gl16}. As the spin of the pre-merging stars has an impact on the properties of the remnant \citep{kast17,east19,papan22}, information gleaned from double precursors may reduce the error bars from afterglow analysis.
It is worth pointing out that a merger that leaves behind a stable NS, rather than a black hole or hypermassive remnant (cf. Fig. \ref{fig:spn}), must be composed of relatively light stars, likely having formed through ``bare collapse'' or electron capture supernovae \citep{pod04}. 

\subsection{Other scenarios leading to double precursors}\label{sec:diff}

Although we focus on $g$-mode scenarios in this work, it is worth briefly commenting on other possibilities, and what one may infer under those circumstances. Indeed, precursors may arise from:
\begin{itemize}
\item[$\star$]{Resonances from other modes, such as $i$- or $s$-modes \citep{Tsang:2011ad,Tsang:2013mca}, ocean modes in low metallicity crusts \citep{sull22}, or even $f$- or $r$-modes in rapidly rotating or ultra-magnetised systems \citep{Suvorov:2020tmk}.}
\item[$\star$]{Each star undergoing a separate fracture, rather than one star undergoing two. Different atomic impurities in the crust, for example, could imply that the von Mises (or some other) criterion is met at different strains and frequencies in each star. Such a scenario would be favoured if, for example, the stellar crust cannot ``heal'' in between fractures; see \cite{schei18,ker22} and references therein.}
\item[$\star$]{Scenarios unrelated to modes, such as the unipolar inductor model, where electromotive forces, generated across a weakly-magnetised star as it moves through the magnetosphere of a magnetar companion, spark precursor emissions \citep{piro12,lai12}. }
\end{itemize}

It is beyond the scope of this work to go into detail about each of these possibilities, though we explore the first point here. In an agnostic analysis, the two precursors of GRB 090510 corresponds to two $l=2=m$ modes (see Sec.~\ref{reso}) with inertial-frame frequencies of $\sim160$ Hz and $\sim50$ Hz, respectively, assuming an aligned system (cf. Sec. \ref{Sec:tilted}).
Given the large difference in the frequencies of $i$- and $s$-modes \cite[see, e.g., Fig.~3 in][]{Passamonti:2011mc}, it seems difficult to connect \emph{both} of these crust-induced modes to the two precursors of GRB090510. In particular, even though the earlier flare may be accommodated by an $i$-mode for an almost-static star, it is difficult to accommodate the later flare with $i$-modes \citep{Tsang:2011ad,Passamonti:2020fur}. 
However, we note that it could be possible to account for these two pre-emissions with a mix of $i$- and $s$-modes, which calls for an $s$-mode with free mode frequency $\sim 160$ Hz since, for an $i$-mode to cause first precursor, we must have $\spin \ll 10$Hz. This mixed-mode scenario will be investigated thoroughly elsewhere.

Some stars in coalescing binaries may spin rapidly, e.g., the secondary of GW190814 \citep{Biswas:2020xna}, where modes with high free frequencies (e.g., $f$-modes) become of interest \citep{Suvorov:2020tmk}. Nonetheless, if such high spins align with the orbit, we should see a rotation-induced modulation in the light-curves during the subsecond timescale of the observed precursors \citep{stach21}. Unless the spin is misaligned with the orbit, the absence of substructure hints that $\spin\lesssim100$ Hz for precursor hosts.

\section{Discussion and summary}
\label{sec:summary}

A system that displays a double precursor, is bright/near enough ($\lesssim 100$~Mpc with aLIGO) to be detected in GWs at merger and late inspiral, and shows an X-ray plateau post-GRB might be something of a holy grail for high-energy astrophysics. As shown here, the first two of these observation bundles may allow for the mass, radius, and spin of (at least one of) the pre-mergering stars to be determined with high accuracy. In principle, this can then be connected to the properties of the post-merger remnant by combining numerical, merger simulations \citep{kast17,east19,papan22} with the spin-down luminosity inferred from the jet energetics \citep{ciolfi20} and X-ray plateau \citep{rowl13}. One may therefore be able to establish a magnetic field strength, compactness, and spin for the remnant \citep{suvk21a,suvk21b,man21}, which imply constraints on the nuclear EOS \cite[e.g.,][]{Biswas:2020xna}. If the {(unstable part of the)} QNM spectrum of the {rapidly spinning} remnant is also observable \cite[out to $\gtrsim 200$ Mpc with the Einstein Telescope; ][]{don15}, the error bars may significantly shrink.

Binary NSs tend to spin slowly; the fastest known binary pulsar is J0737-3039A with a spin frequency of 44 Hz \citep{Burgay:2003jj}. For GRB 090510 we predict that the star emitting the two precursors has $\spin\approx 3$~Hz if $l=2=m$ $g_1$- and $g_2$-mode resonances spark the precursors (Tab.~\ref{tab:m1m2}), which is lower than this value. Note that the spin estimate given above is insensitive to the EOS {(modulo the caveats mentioned in Sec. \ref{reso}) and is also insensitive to $q$ (Fig.~\ref{fig:nu_q})}.
On the other hand, if the associated modes have different combinations of quantum numbers because of a significant tilt, $\Theta \gtrsim 30^{\circ}$ \cite[on par with the greatest limit set on PSR J1756-2251, $\leq 34^\circ$, as inferred from geodetic precession;][]{ferd14}, the inferred spin ranges from $2 \lesssim \nu_{\star}/\text{Hz} \lesssim 13$ assuming instantaneous jet breakout and perfect precision on the timing measurements. In reality, the uncertainty is more likely to cover the range $2 \lesssim \nu_{\star}/ \text{Hz} \lesssim 20$ if we remain agnostic regarding these factors (see Tab.~\ref{tab:m1m2} and Sec. \ref{sec:jet}). {This is especially true because of our simplified prescription for introducing the adiabatic index of the perturbed star; see the caveats noted in Sec. \ref{reso} and throughout.} Despite the uncertainty in the quantum numbers, we only have four possibilities for the spin, any of which can serve as a necessary condition when accounting for double precursors via $g_1$- and $g_2$-modes (see the discussion in Sec.~\ref{Sec:spinfreqdet}), and the method demonstrates, in principle, how additional information can be extracted from doubles.

In addition to frequency matching, the modes must be excited strongly enough to shatter the crust. The extent to which a mode can be amplified by the tidal field depends on the mass of the other star, the inclination of the spin, and the tidal overlap of the mode. In other words, an observation of two precursors can place certain constraints on the range of the aforementioned quantities. For example, a star with mass $1.45M_{\odot}$ may not able to yield its crust via $g$-mode resonances due to the tidal neutrality of these modes \citep{prep}. 
In addition, non-zero inclinations will weaken the mode excitations in general (not just $g$-modes), thus the observation of resonant shattering flares can set a limit on the tilt angle of erupting NSs. Together with the inferred spin, such a study may shed light on the binary formation channel; for example, a NS member spinning at a low rate with a small tilt angle can likely exclude a scenario where the primary is long-term recycled and aligned (see Sec.~\ref{sec:form} for more discussion).

Although only one double precursor event (in a short GRB) has thus far been observed, in the future one may be able to -- assuming a resonance scenario -- put statistical constraints on the spin dynamics of NS binaries that do not exhibit pulsations. This allows for an investigation of the evolutionary pathways of NSs that reside within the so-called pulsar `graveyard'. It is conceivable also that even millisecond objects may enter into compact binaries in dense astrophysical environments through dynamical exchanges. What might the timing data of double precursors look like in such a case? For spin frequencies $\nu_{\star} \gg 10$Hz, equation \eqref{eq:spin} implies that the two events should be separated by at least 15 seconds, a prediction which is robust for different EOS. Depending on the spin alignment of the binary constituents, large values of $\spin$ may excite the so-called ``one-arm'' instability in the GW spectrum of the remnant and enhance blue/red kilonovae \citep{east19,papan22}.

We close by noting that magnetic fields have not been considered at all in this work. It is likely that magnetic fields play a significant role in extracting the elastic energy from the crust that eventually fuels the precursor \citep{Tsang:2013mca,Suvorov:2020tmk,suvo22}. However, unless the fields are of magnetar-level strength ($B_{\star} \gtrsim 10^{15}$~G), the Lorentz force will not be strong enough to significantly distort the QNM spectrum \citep{Kuan:2021jmk}, implying that the spin-fitting formula \eqref{eq:spin} would remain unchanged. Even so, \cite{Suvorov:2020tmk} argued that precursors with non-thermal spectra may be indicative of intense magnetic fields, so as to avoid thermalisations from pair-photon cascades created via  mode-induced backreactions \cite[see also][]{Tsang:2011ad,Zhong:2019shb,Kuan:2021sin}. These considerations imply that the error bars presented on the spin-frequency measurements may be slightly underestimated therefore, at least when applied to double precursors showing  predominantly non-thermal spectra \cite[as indeed was the case for GRB090510;][]{Troja:2010zm}. 

\begin{acknowledgements}
KK gratefully acknowledges financial support by DFG research Grant No. 413873357. HJK recognises support from Sandwich grant (JYP) No. 109-2927-I- 007-503 by DAAD and MOST during early stages of this work. AGS is grateful for funding received from the
European Union's Horizon 2020 Programme under the AHEAD2020 project (grant n. 871158) and the Alexander von Humboldt foundation. 
We thank the anonymous referee for their helpful feedback.
\end{acknowledgements}

\bibliographystyle{aa}
\bibliography{references}

\appendix
\section{Computation of $g$-mode frequencies with spatially-dependent stratification} \label{sec:app}

This Appendix details the $g$-mode frequencies for two different cases: one where we fix $\delta$ to a (well-motivated) constant parameter, and one where we instead consider an isothermal star and compute $\delta$ self-consistently via expression \eqref{eq:deltadef}. As noted in the main text, in estimating $T$ from \eqref{eq:deltadef} we assume that thermal pressure is dominated by non-relativistic $n$ and $p$, whose Fermi energies are given by \cite{kruger15}
\begin{align}\label{eq:ef}
    E_F^x(\boldsymbol{x})=\frac{\hbar^2[3\pi^2n_x(\boldsymbol{x})]^{2/3}}{2m_x^*},
\end{align}
with $n_x$ and $m_x^*$ denoting, respectively, the number density and the Landau effective mass of the species, and further that the effective masses of $n$ and $p$ coincide and are approximated as $m_n^*=m_p^*=0.8$ times of nucleon mass following \cite{cham06}.

\begin{figure}
	\centering
	\includegraphics[width=0.497\textwidth]{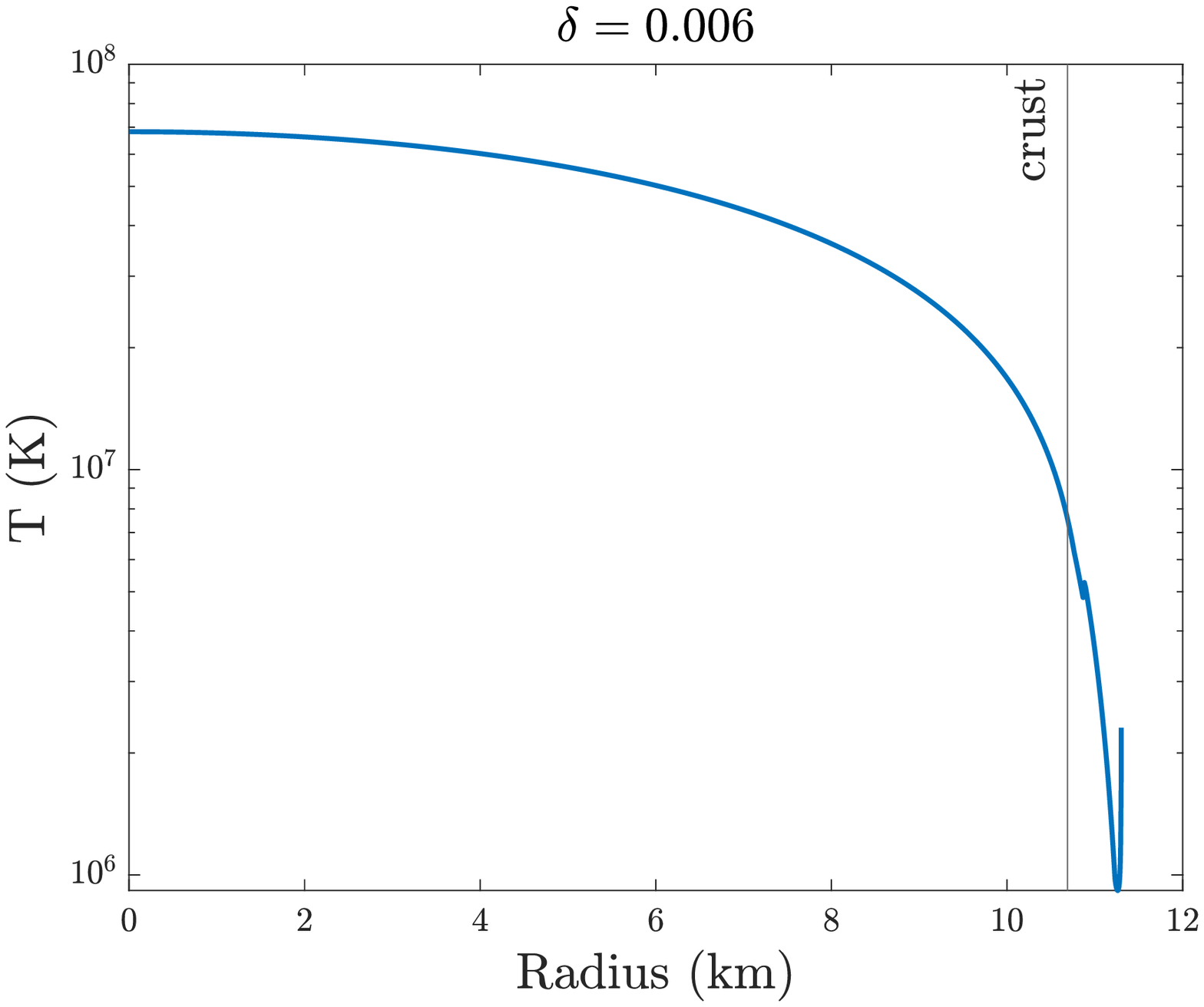}
	\includegraphics[width=0.497\textwidth]{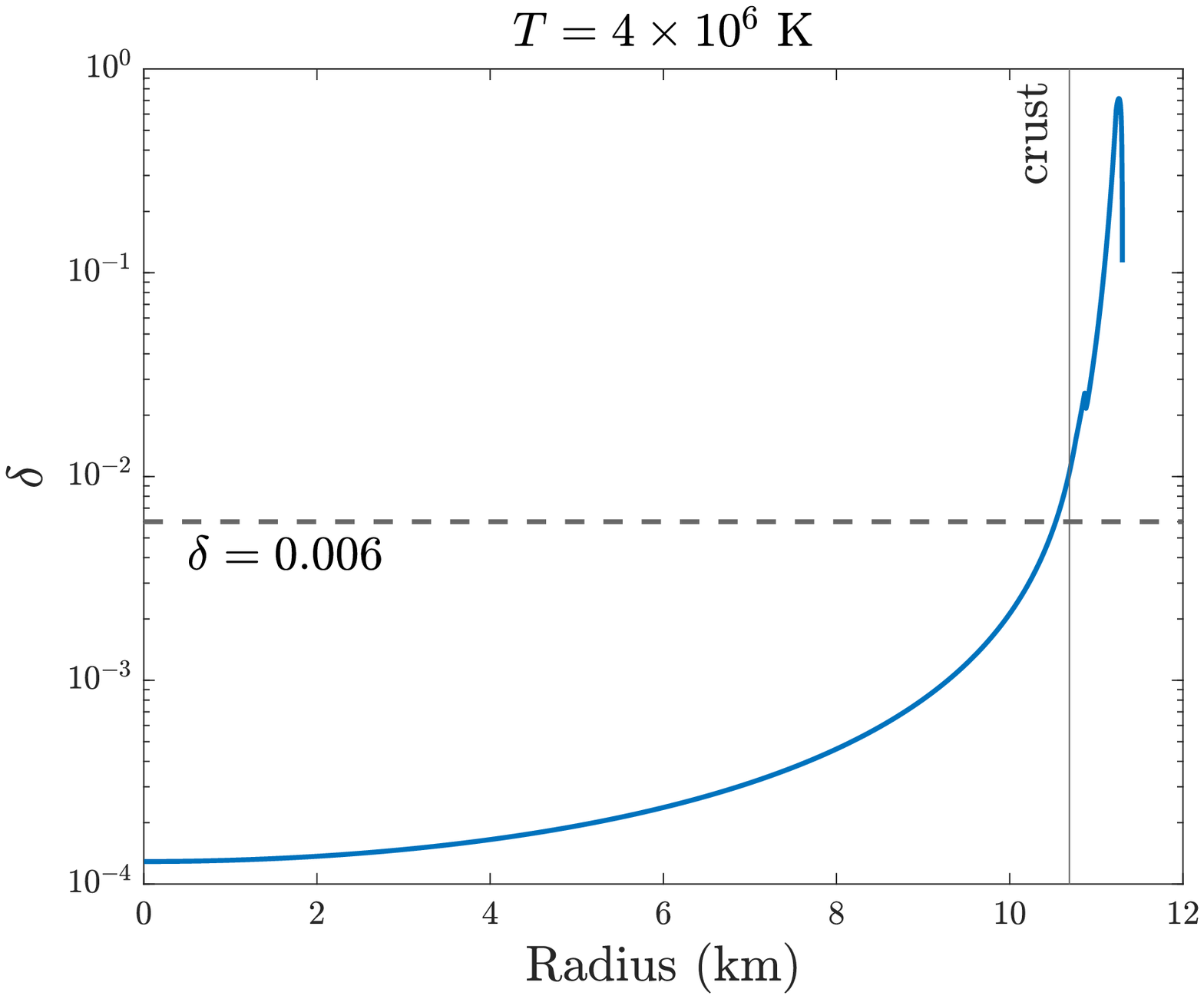}
	\caption{\textit{top}: profile of $T$ for a star with constant $\delta = 0.006$. The crust-core boundary is indicated. \textit{bottom}: profile of $\delta$ for constant temperature, where the dashed line represents the value that we use for the first precursor of GRB090510 (see the main text). For both panels, the equations \eqref{eq:deltadef} and \eqref{eq:ef} are used, and a NS pertaining to EOS SLy4 with a mass of 1.41 $M_\odot$ is adopted.}
	\label{fig:def}
\end{figure}

Figure~\ref{fig:def} shows the implied temperature profile for a constant $\delta$ (top) and, by contrast, the inferred $\delta$ for an isothermal star with the same surface temperature (bottom), both with a SLy4 EOS and $M_{\star}=1.41 M_\odot$. The $g_1$- and $g_2$-mode frequencies are computed as ($f_{g_1},f_{g_2}$)=(101.10 Hz, 70.48 Hz) for the left case and ($f_{g_1},f_{g_2}$)=(101.67 Hz, 66.99 Hz) for the right case. Interestingly, the chosen temperature in the right case is roughly the \emph{volume-average} over the \emph{crust} region of the former case. We see that the frequencies of $g_1$- and $g_2$-modes deviate by only $\lesssim5\%$ between the two cases, meaning that a constant-$\delta$ approximation works well, despite this parameter varying by several orders of magnitude, $10^{-4} \lesssim \delta \lesssim 1$, within the stellar interior: it is mostly the crust-core transition value that is important. More complicated and physically motivated $T$ profiles were also investigated while preparing  \cite{prep} (though not shown explicitly), validating this assumption more generally. We note that we adopt the SLy EOS in this Appendix since the compositional particulars can be found in table format within the freely-accessible CompOSE catalogue \citep{type15,oert17,type22}. In particular, the neutron and proton fractions are given as a function of energy density, which then determine the relation between $\delta$ and $T$ at the some time $t$ and position $\boldsymbol{x}$ inside the star.

\label{lastpage}
\end{document}